\documentclass{aa}  

\usepackage{hyperref}
\usepackage{color}
\usepackage{graphicx}
\usepackage{txfonts}
\usepackage{newtxtext,newtxmath}
\DeclareSymbolFont{CMletters}{OML}{cmm}{m}{it}
\DeclareMathSymbol{v}{\mathord}{CMletters}{`v}

\usepackage{adjustbox}

\newcommand{\kms}{km\,s$^{-1}$}
\newcommand{\ms}{m\,s$^{-1}$}
\newcommand{\vsini}{$v \sin i_\ast$ }
\newcommand{\HTWOO}{$\textrm{H}_2 \textrm{O}$ }

\newcommand{\M}{MASCARA-4\,b }

\usepackage{silence}
\begin{document} 

   \title{Transmission spectroscopy of the ultra-hot Jupiter \M \thanks{Transit photometry and radial velocity data are only available in electronic form at the CDS via anonymous ftp to \href{ftp://130.79.128.5}{cdsarc.u-strasbg.fr (130.79.128.5)} or via \url{http://cdsweb.u-strasbg.fr/cgi-bin/qcat?J/A+A/}}}
    \subtitle{Disentangling the hydrostatic and exospheric regimes of ultra-hot Jupiters}
    
    \author{Yapeng Zhang \inst{1}, Ignas A.G. Snellen \inst{1},  Aur\'elien Wyttenbach \inst{2}, Louise D. Nielsen \inst{3,2}, Monika Lendl \inst{2}, N\'uria Casasayas-Barris \inst{1},  Guillaume Chaverot \inst{2}, Aurora Y. Kesseli \inst{5}, Christophe Lovis \inst{2}, Francesco A. Pepe\inst{2},  Angelica Psaridi \inst{2}, Julia V. Seidel \inst{2, 4}, St\'ephane Udry \inst{2}, Sol{\`e}ne Ulmer-Moll \inst{2}
    }
   \institute{Leiden Observatory, Leiden University, Postbus 9513, 2300 RA, Leiden, The Netherlands \\
   \email{yzhang@strw.leidenuniv.nl}
   \and 
   Observatoire astronomique de l'Universit{\'e} de Gen{\`e}ve, Chemin Pegasi 51, 1290 Versoix, Switzerland
   \and
   European Southern Observatory, Karl-Schwarzschildstr. 2, D-85748 Garching bei M{\"u}nchen, Germany 
  \and
  European Southern Observatory, Alonso de C\'ordova 3107, Vitacura, Regi\'on Metropolitana, Chile
  \and
  IPAC, Mail Code 100-22, Caltech, 1200 E. California Blvd., Pasadena, CA 91125, USA
   }
    \authorrunning{Y. Zhang et al.}
    \titlerunning{MASCARA-4b}
    
  \date{Received June 7, 2022; accepted August 16, 2022}

\abstract{
Ultra-hot Jupiters (UHJs), rendering the hottest planetary atmospheres, offer great opportunities of detailed characterisation with high-resolution spectroscopy.  \M is a recently discovered close-in gas giant belonging to this category.
}
{
We aim to characterise MASCARA-4~b, search for chemical species in its atmosphere, and put these in the context of the growing knowledge on the atmospheric properties of UHJs.
}
{
In order to refine system and planet parameters, we carried out radial velocity measurements and transit photometry with the CORALIE spectrograph and EulerCam at the Swiss 1.2\,m Euler telescope. 
We observed two transits of \M with the high-resolution spectrograph ESPRESSO at ESO’s Very Large Telescope.
We searched for atomic, ionic, and molecular species via individual absorption lines and cross-correlation techniques. These results are compared to literature studies on UHJs characterised to date.
}
{
With CORALIE and EulerCam observations, we update the mass of \M ($M_{\rm p}=1.675\pm0.241$ $M_{\rm Jup}$) as well as other system and planet parameters.
In the transmission spectrum derived from ESPRESSO observations, we resolve excess absorption by H$\alpha$, H$\beta$, \ion{Na}{i} D1\&D2, \ion{Ca}{ii} H\&K, and a few strong lines of \ion{Mg}{i}, \ion{Fe}{i}, and \ion{Fe}{ii}. We also present the cross-correlation detection of \ion{Mg}{i}, \ion{Ca}{i}, \ion{Cr}{i}, \ion{Fe}{i}, and \ion{Fe}{ii}. 
The absorption strength of \ion{Fe}{ii} significantly exceeds the prediction from a hydrostatic atmospheric model, as commonly observed in other UHJs. 
We attribute this to the presence of \ion{Fe}{ii} in the exosphere due to hydrodynamic outflows. This is further supported by the positive correlation of absorption strengths of \ion{Fe}{ii} with the H$\alpha$ line, which is expected to probe the extended upper atmosphere and the mass loss process.
Comparing transmission signatures of various species in the UHJ population allows us to disentangle the hydrostatic regime (as traced via the absorption by \ion{Mg}{i} and \ion{Fe}{i}) from the exospheres (as probed by H$\alpha$ and \ion{Fe}{ii}) of the strongly irradiated atmospheres. 
}
{} 


   \maketitle
%

\section{Introduction}

Transmission spectroscopy of close-in giant planets provides great opportunities to characterise the composition, structure, and dynamics of exoplanet atmospheres \citep{Charbonneau2002, Redfield2008, Snellen2010, Huitson2012, Madhusudhan2019}.
These strongly irradiated planets undergo significant atmospheric escape as traced by absorption signatures from exospheres extending beyond the Roche limit \citep{Vidal-Madjar2003, Vidal-Madjar2004, Spake2018}. The mass loss drives the evolution of close-in planets and shapes the exoplanet population as observed today \citep{Owen2019}. 

Ultra-hot Jupiters (UHJs) represent a subclass of close-in hot Jupiters that are extremely irradiated, with day-side temperatures above 2200 K. As a result of such high temperatures, their day-side atmospheres are predicted to be cloud-free and with effective thermal dissociation of molecules such as \HTWOO\ to produce OH \citep{Parmentier2018, Nugroho2021, Landman2021}. The transmission spectra are dominated by neutral and ionized atomic species in the optical, similar to the photosphere of dwarf stars \citep{Kitzmann2018, Lothringer2018}, which are well suited for atmospheric characterisation. The dissociation of hydrogen combined with electrons from metal ionisation to form H$^-$, adds strong continuum opacity which plays an important role in shaping the spectra \citep{Arcangeli2018}.   
The extreme irradiation also makes UHJs interesting targets for studying the mass loss via hydrodynamic escapes \citep{Fossati2018, Sing2019}.

Transmission spectra under high spectral resolution ($\mathcal{R}=\lambda/\Delta\lambda\sim10^5$) provide unique access to the information contained in resolved absorption lines, such as \ion{Na}{i} D lines, \ion{H}{i} Balmer series, and \ion{He}{i} triplet, allowing better constraints on the atmospheric structure and the escaping process \citep[e.g.][]{Wyttenbach2015, Wyttenbach2017, Yan2018, Allart2018, Nortmann2018, Casasayas-Barris2019, Wyttenbach2020}.  
In addition, high-resolution spectroscopy has been a powerful tool that leads to the detection of a profusion of metal species in UHJs using the cross-correlation method that co-adds a forest of spectral lines to enhance the signal of a certain species \citep{Snellen2010, Brogi2012}. For example, \citet{Hoeijmakers2018, Hoeijmakers2019} detected neutral and ionized metals (such as \ion{Fe}{i}, \ion{Fe}{ii}, \ion{Ti}{ii}, \ion{Cr}{ii}, etc.) in KELT-9b, the hottest known planet \citep[T$_\mathrm{eq} $$\sim$4000 K,][]{Gaudi2017}. Subsequent studies have quickly extended detections to more UHJs, including  WASP-121b \citep{Ben-Yami2020, Gibson2020, Hoeijmakers2020a, Merritt2021}, MASCARA-2b \citep{Stangret2020, Nugroho2020, Hoeijmakers2020b}, WASP-76b \citep{Kesseli2022}, TOI-1518b \citep{Cabot2021}, WASP-189b \citep{Prinoth2022}, and HAT-P-70b \citep{Bello-Arufe2022}.

Here we present the transmission spectroscopy of \M \citep{Dorval2020}, an ultra-hot Jupiter with an equilibrium temperature of $\sim$2250 K, orbiting at 0.047 au away from an A7V star ($m_\mathrm{V}=8.2$) with an effective temperature of $\sim$7800 K. The properties of the system are summarised in Table~\ref{tab:M4}.
Two transit observations were taken with Echelle SPectrograph for Rocky Exoplanets and Stable Spectroscopic Observations \citep[ESPRESSO,][]{Pepe2021} at the VLT. 
This analysis adds \M to the ensemble of UHJs that have been characterised with high-resolution transmission spectroscopy and show absorption features from various atomic species.

We describe the observations and data reduction in Section~\ref{sec:observation}. The analyses are presented in Section~\ref{sec:data-analysis}, including fitting the spin-orbit misalignment angle, modeling the Rossiter-McLaughlin (RM) effect, extracting transmission spectra, and carrying out cross-correlation. We then present in Section~\ref{sec:result} the detection of planetary absorption signals in the transmission spectrum from both single-line and cross-correlation analysis. In Section~\ref{sec:discussion}, we put the results of \M in context of the UHJ population and discuss trends of absorption strengths among UHJs that may shed light on the atmospheric structures. 

\begin{table}[]
\centering
\caption{Properties of the MASCARA-4 system.}
\resizebox{\columnwidth}{!}{
\begin{tabular}{lc}
\hline \hline
 Parameter  & Value \\ 
 \hline
 \multicolumn{2}{c}{\dotfill\it Stellar parameters\dotfill}\\\noalign{\smallskip}
Effective temperature, $T_{\rm eff}$ (K) $^1$ & $7800 \pm 200$ \\
Stellar mass, $M_\ast$ ($M_\odot$) $^1$ & $1.75\pm0.05$ \\
Stellar radius, $R_\ast$ ($R_\odot$) $^2$ & $1.79\pm0.04$ \\
Surface gravity, log($g$) $^1$ & $4.10\pm0.05$ \\
Projected spin, \vsini (\kms) & $43.0\pm0.1$ (Spectral)\\
 &  $46.2^{+7.7}_{-2.5}$ (RM reloaded)  \\
Differential rotation, $\alpha$ & $0.09\pm0.03$ (RM reloaded) \\
 Limb-darkening coeff., $u_1$  & 0.333 \\
Limb-darkening coeff., $u_2$ & 0.332 \\  
\hline
 \multicolumn{2}{c}{\dotfill\it Updated system parameters \dotfill}\\\noalign{\smallskip}
 RV amplitude, $K_{\ast}$ (\ms) & $165.9\pm 23.7$ \\
 Mid-transit time, $T_0$ (BJD) & $ 2458909.66419\pm 0.00046  $  \\
 Transit duration, $T_{14}$ (day)  &  $  0.1654 \pm 0.0013 $   \\
 Orbital period, $P$ (day) &  $2.8240932 \pm 0.0000046 $ \\
 Radius ratio,  $R_{\rm p}/R_\ast$   &  $ 0.0869 \pm 0.0015 $   \\
 Impact parameter, $b$  &   $  0.309 \pm 0.044 $     \\
 Orbital inclination , $i_{\rm p}$ (deg)    &    $    86.89 \pm 0.49 $          \\
 Semi-major axis, $a$ (au)               &       $  0.0474 \pm 0.0013 $      \\
 $a/R_\ast$                  &       $  5.704_{-0.096}^{+0.086} $ \\  
 Spin-orbit angle, $\lambda$ (deg) & $250.34\pm0.14$  \\
 \hline
 \multicolumn{2}{c}{\dotfill\it Updated planet parameters \dotfill}\\\noalign{\smallskip}
Planetary radius,  $R_{\rm p}$ ($R_{\rm Jup}$)  & $ 1.515 \pm 0.044 $  \\
Planetary mass, $M_{\rm p}$ ($M_{\rm Jup}$) & $1.675\pm0.241$ \\
Planetary density, $\rho_{\rm p}$ ($\rho_{\rm Jup}$)       &      $   0.481_{-0.079}^{+0.085}  $ \\
Surface gravity, $g_{\rm p}$ (m\,s$^{-2}$)          &       $ 18.1 \pm 2.9 $            \\
Equilibrium temperature, $T_{\rm eq}$ (K)  & $2250 \pm 62$ \\
Semi-amplitude velocity, $K_{\rm p}$ (\kms) & $182\pm 5$ \\
\hline
\end{tabular}
}
\tablebib{ $(1)$ \citet{Dorval2020}; $(2)$ \citet{Ahlers2020}.}
\label{tab:M4}
\end{table}


\section{Observations and data reduction} \label{sec:observation}

\subsection{Radial velocity measurements with CORALIE and updated planet mass} \label{sec:coralie}

   \begin{figure}
   \centering
   \includegraphics[width=\hsize]{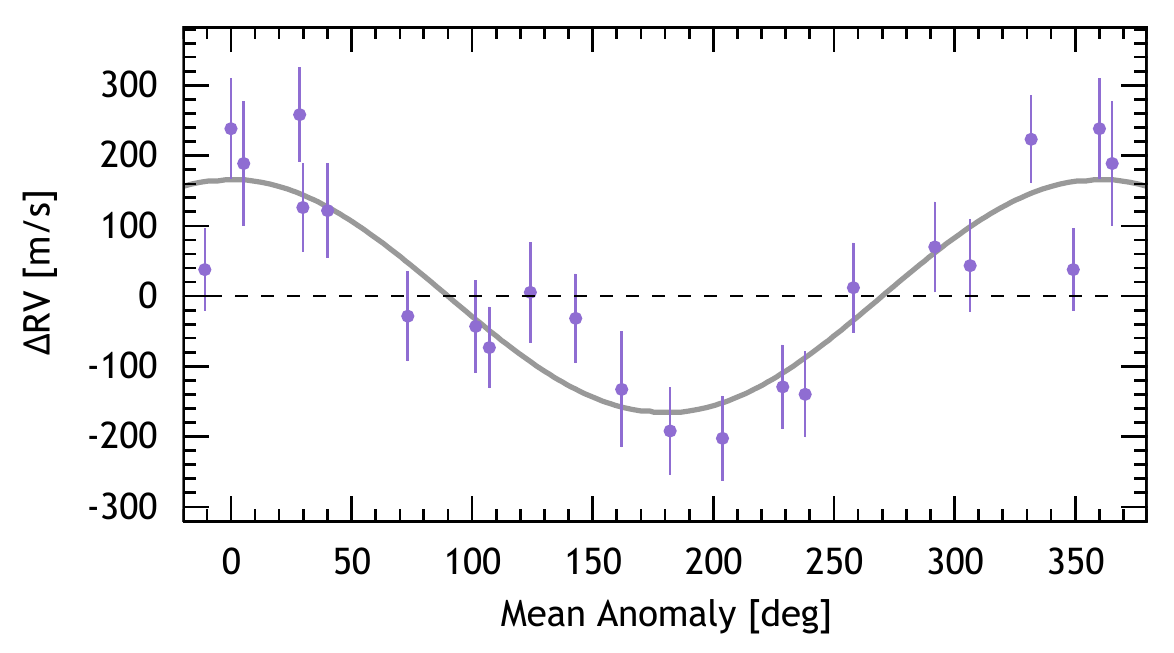}
      \caption{CORALIE radial velocity measurements of MASCARA-4, phase folded on the planetary orbital period and corrected for steller activity. The best-fit Keplerian model to the data is shown as a solid grey line. The semi-amplitude of the stellar reflex motion is $165.9\pm 23.7$ \ms, corresponding to a planet mass of $1.675\pm0.241$ $M_{\rm Jup}$.
              }
         \label{fig:rv}
   \end{figure}
   
To refine the mass of \M, 20 high-resolution spectra were obtained with the CORALIE spectrograph \citep{Baranne1996, Queloz2000} at the Swiss 1.2\,m Euler telescope at La Silla Observatories, Chile. The observations took place between January 4, 2020 and February 10, 2021. The SNR per pixel at 550 nm varied between 55 and 90, according to sky condition and exposure time (20-30 min depending on target visibility and scheduling requirements). 

Radial velocity (RV) measurements were extracted by cross-correlating the spectra with a binary A0 mask. The shape of the cross-correlation functions (CCFs) were dominated by rotational broadening (FWHM $\approx 55$ \kms) and a linear slope across the continuum due to imperfect flux correction across the CCF window. Similar to the approach demonstrated for WASP-189 \citep{Anderson2018}, we fitted a rotational profile with a linear slope to the CCFs \citep[for more information, see Sect. 2.3.3 in][]{NielsenThesis}. The inverse bisector-span \citep[BIS,][]{Queloz2001} was computed on the continuum-corrected CCFs.  

The RV fitting were carried out with the Data and Analysis Center for Exoplanets web platform (DACE) \footnote{\url{https://dace.unige.ch/}}. The Keplerian model described in \citet{Delisle2016} was fit to the RV data points using a Markov chain Monte Carlo (MCMC) algorithm \citep{Diaz2014, Diaz2016}, while applying Gaussian priors on the stellar mass and planetary orbit ($P$, $T_0$ and $i_p$) from Table \ref{tab:M4}. When allowing for an eccentric orbit, we found an eccentricity consistent with zero and adopted a circular model to avoid overestimating the eccentricity \citep{2019MNRAS.489..738H}. A traditional RV fit yielded a moderate anti-correlation between the RV-residuals and BIS (weighted Pearson coefficient $R_w = -0.68$). We, therefore, applied a linear detrending of the RVs with BIS in the model. 
The final RV analysis derives a semi-amplitude of the stellar reflex motion of $165.9\pm 23.7$ \ms, corresponding to a planet mass of $1.675\pm0.241$ $M_{\rm Jup}$. We tested the fitting without the detrending step and found no sensible change in the resultant semi-amplitude.
The phase folded data, detrended with BIS, is shown in Fig.~\ref{fig:rv} along with the Keplerian model. 
The planet mass deviates by $2\sigma$ from the previous measurement of $3.1\pm0.9$ $M_{\rm Jup}$ (and RV semi-amplitude $310 \pm 90$ \ms) in \citet{Dorval2020}, which relied on one particular data point with large uncertainty. Hence we adopt the revised values.

\subsection{Photometry with EulerCam}
\label{sec:ecam}

\begin{figure}
\centering
\includegraphics[width=\linewidth]{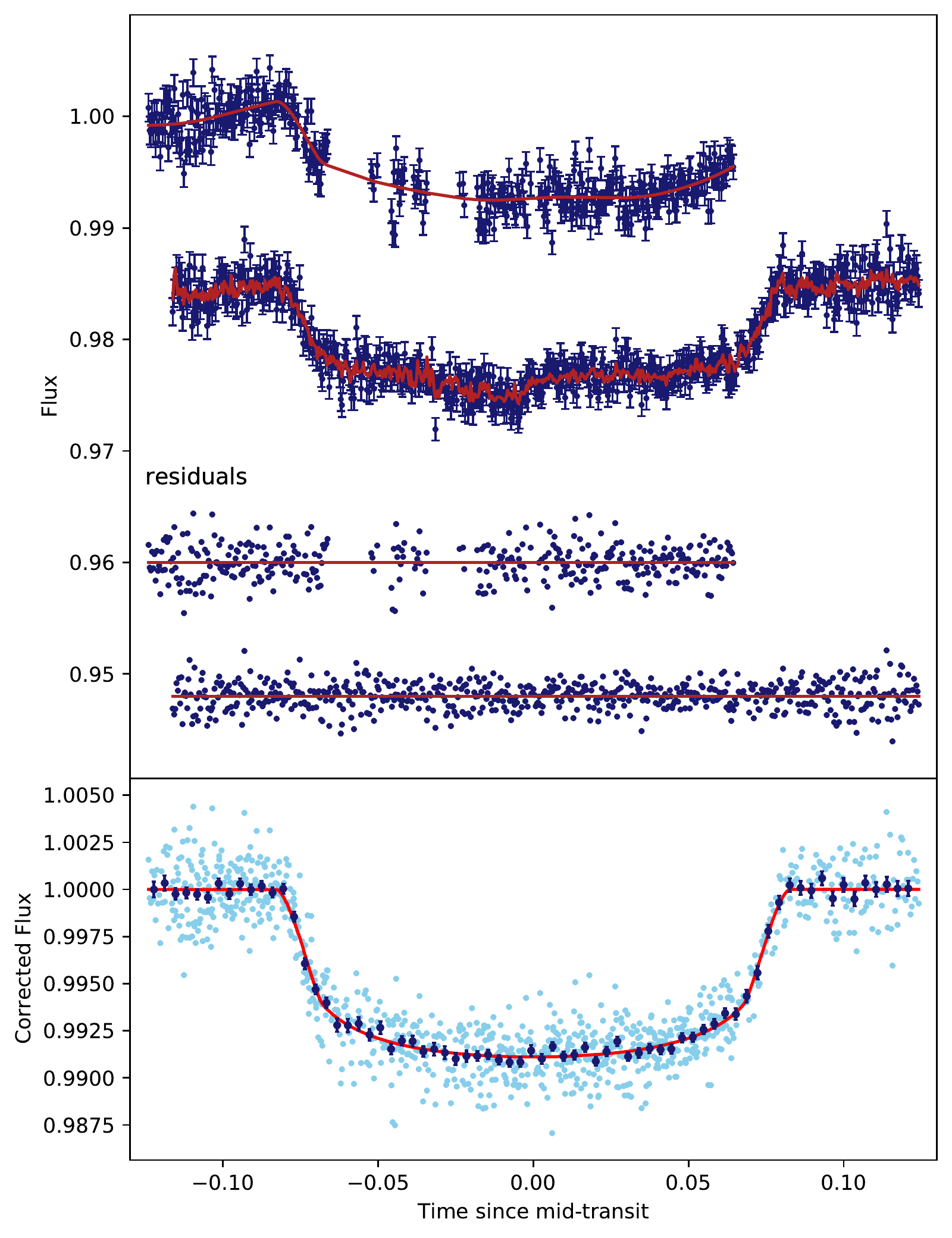}
    \caption{\label{fig:LC}EulerCam light curves of MASCARA-4. \emph{Top panel:} raw differential photometry together with the transit and systematic model corresponding to the median posterior values. The top two light curves are the data, while the bottom two light curves show the residuals to these models.
    \emph{Bottom panel:} Systematics-corrected, phase-folded data together with the transit model. The unbinned data points are shown in light blue, while the dark blue points show the data binned into 2-minute intervals.}
       \label{FigVibStab}
  \end{figure}

We observed two transits of \M on February 12 and 29, 2020 using EulerCam, the CCD imager installed at the 1.2~m Euler telescope located at La Silla. The observations were scheduled to be simultaneous with the two nights of observations with ESPRESSO (see Section~\ref{sec:espresso}), delivering updated transit parameters for the analysis of transmission spectroscopic data. For more details on the instrument and associated data reduction procedures the reader is referred to \citet{Lendl2012}. 
As the star is brighter (V=8.19) than exoplanet hosts usually observed with EulerCam \citep[V $\sim10-14$,][]{Lendl2012, Lendl2013, Lendl2019}, using a broad-band filter would lead to saturation of the detector. We therefore used a narrower band to avoid saturation issues, namely the \emph{Geneva V1} filter \citep{Rufener1988}, which peaks at 539\,nm and has a transmission above 50\% from 509\,nm to 562\,nm (accounting for detector quantum efficiency). The telescope was also defocused slightly to improve PSF sampling and observation efficiency. An exposure time of 20~s was used throughout both sequences. The light curves shown in Fig. \ref{fig:LC} were obtained using relative aperture photometry with two bright reference stars and apertures of 26 pixel (5.6\arcsec) radius. The night of February 12, 2020 was affected by recurrent cloud passages, leading to gaps in the observed data.

We used the EulerCam data to compute the physical system parameters and in particular derive a planetary radius in the \emph{Geneva V1} band, which is comparably close in wavelength covered with ESPRESSO. TESS \citep{Ricker2015} has previously observed \M, revealing a slightly asymmetric transit shape created by gravity darkening on the host star and a misaligned planetary orbit \citep{Ahlers2020}. Our ground-based observations do not possess sufficient precision to reveal this effect. However, to propagate the information encoded in the TESS data into our fit, we place Gaussian priors on the transit duration ($T_{\mathrm{14}}$) and the impact parameter ($b$) corresponding to the values presented by \citet{Ahlers2020}. We used a Markov Chain Monte Carlo approach as implemented in \emph{CONAN} \citep{Lendl2020b} to derive the system parameters, fitting for  $R_p / R_\ast$, $b$, $T_{\mathrm{14}}$, $T_0$ and $P$. With the exception of $b$ and $T_{\mathrm{14}}$ for which broad uniform priors were assumed. We assumed a quadratic limb-darkening law with parameters derived with LDCU\footnote{\url{https://github.com/delinea/LDCU}} \citep{Deline2022}. Correlated noise was modelled individually for each light curve using approximate Mat\'ern-3/2 kernels implemented through \emph{celerite} \citep{FM2017}. For the light curve of February 12, 2020, we included an evident correlation between the residual flux and the stellar FWHM as a linear trend fitted together with the transit model and Gaussian Process (GP). We allowed for an additional white noise by inclusion of a jitter term for each light curve.

To derive planetary parameters, we used our derived radial velocity amplitude of $165\pm23$~\ms as presented in Section~\ref{sec:coralie} and pulled values from a corresponding Normal distribution at each MCMC step. Similarly, normal distributions for $M_\ast$ and $R_\ast$ were assumed corresponding to the values inferred from our spectral analysis. The raw and phase-folded light curves are shown in Fig.~\ref{fig:LC}, and the resulting updated parameters are given in Table \ref{tab:M4}.

\subsection{High-resolution transmission spectroscopy with ESPRESSO} \label{sec:espresso}

\begin{table*}
\caption{Observing log of \M with ESPRESSO}
\label{tab:obs}      
\centering                         
\begin{tabular}{c c c c c c c c}     
\hline\hline                
Night & Date & Exposure time (sec) & N$_\mathrm{spectra}$ & On-target time (hour) & airmass & seeing (\arcsec) & S/N@580\,nm \\    
\hline
1 & Feb 12, 2020 & 360 & 85 & 7.5 & 1.34 - 1.99 & 0.34 - 0.87 & $\sim$212\\
2 & Feb 29, 2020 & 300 & 96 & 7.9 & 1.34 - 2.46 & 0.32 - 0.77 & $\sim$208\\
\hline                                  
\end{tabular}
\end{table*}  

We observed two transits of \M with ESPRESSO on February 12 and 29, 2020 under ESO program 0104.C-0605 (PI: Wyttenbach). ESPRESSO is a fiber-fed, ultra-stabilized echelle high-resolution spectrograph installed at the incoherent combined Coud\'e facility of the VLT. The observations were taken with the single-UT HR21 mode, providing a spectral resolving power of $\mathcal{R}\sim138\,000$, covering the optical wavelength range of 380-788 nm. The observations are summarised in Table~\ref{tab:obs}.
The exposure time during transit was 360s and 300s in night 1 and night 2 respectively. The airmass ranges from 1.34 to 2.46, and the seeing condition varied from 0.3\arcsec\ to 0.9\arcsec. The total on-target time is 7.5h (85 exposures) and 7.9h (96 exposures) in the two nights respectively, delivering an average S/N per pixel of 208 and 212 at 580 nm.

We took the sky-subtracted 1D spectra extracted with the Data Reduction Software (DRS) pipeline, and then corrected for telluric absorption features caused by H$_2$O and O$_2$ in the Earth's atmosphere following \citet{Allart2017} using the ESO sky tool \texttt{molecfit} \citep[version 4.2,][]{Smette2015}. The tool uses a line-by-line radiative transfer model (\texttt{LBLRTM}) to derive telluric atmospheric transmission spectra and accounts for molecular abundances, instrument resolution, continuum level, and wavelength solution that can best fit observations, whereas other telluric or interstellar contamination such as the absorption of \ion{Na}{i} was not removed with this correction. 

We also used the ESPRESSO DRS to generate stellar cross-correlation functions (CCFs) with an A0 mask as presented in \citet{Wyttenbach2020}. 
The stellar CCFs outside of the transit were average-combined to build a master out-of-transit CCF, representing the unocculted stellar line shape. We measured the projected spin velocity \vsini and systemic velocity $V_\mathrm{sys}$ of the target by fitting a rotationally broadened model \citep{Gray2005} to the line shape. We found the \vsini of 43.0$\pm$0.1 \kms and the $V_\mathrm{sys}$ of -5.68$\pm$0.09 \kms.
We then obtained the residual CCFs by subtracting the CCF at each phase from the master out-of-transit CCF. The residual CCFs were later used to extract Rossiter-McLaughlin information as detailed in Section~\ref{sec:RM-reloaded}.

\section{Data analysis} \label{sec:data-analysis}

\subsection{Stellar pulsations} \label{sec:pulsation}

\begin{figure}
   \centering
   \includegraphics[width=\hsize]{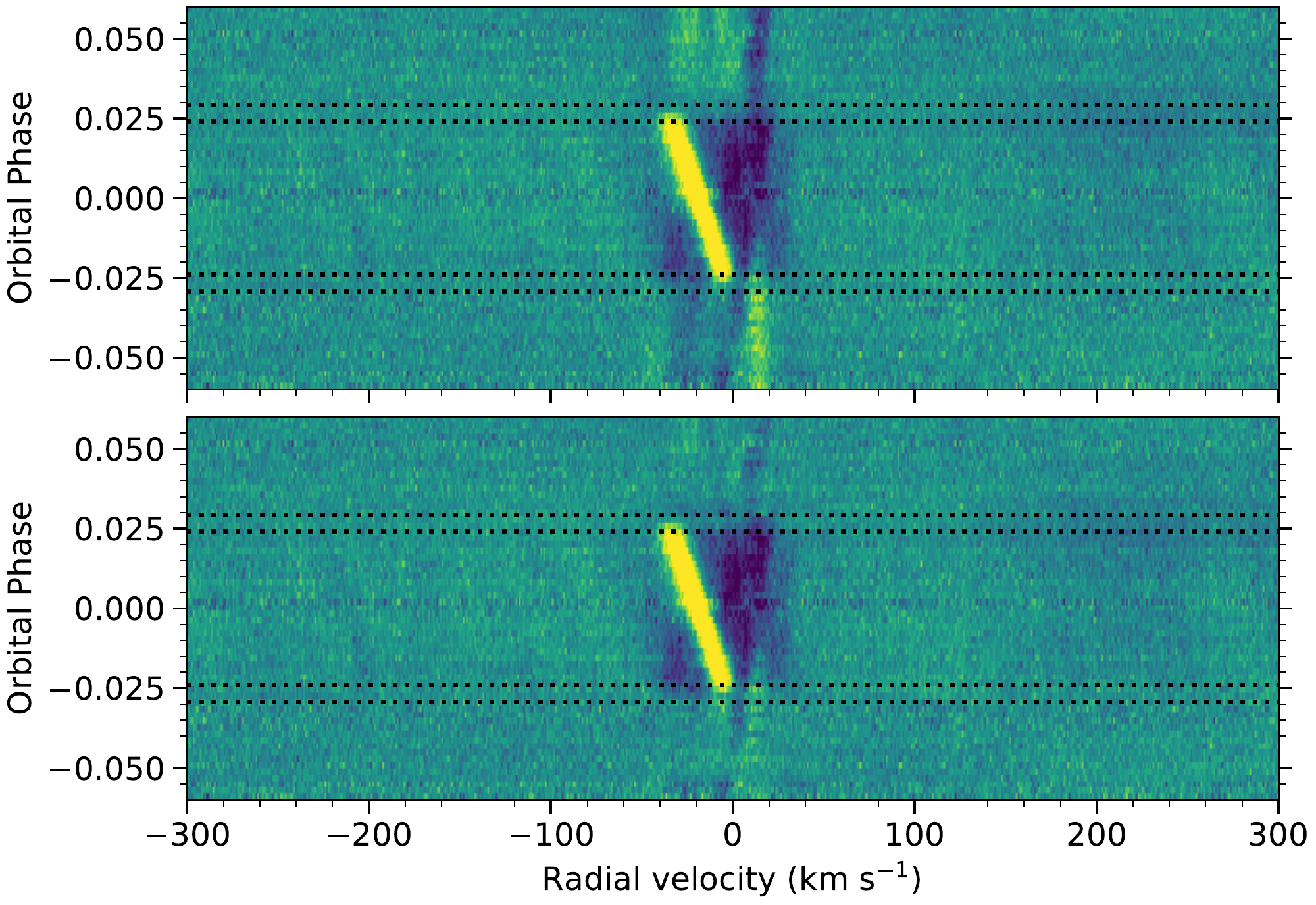}
      \caption{Map of the residual CCFs at different orbital phases before (top panel) and after (bottom panel) the correction of stellar pulsation. Data of two transits were binned in orbital phase with a step size of 0.002. The horizontal dotted lines denote the timings of four transit contacts. The slanted yellow streak during transit delineates the deformation of stellar lines as the planet moving across the stellar disk. The ripples seen in the out-of-transit residual CCFs in the top panel are attributed to stellar pulsation.
              }
         \label{fig:M4_pulsation}
   \end{figure}

In residual CCFs we note a strong rippled pattern caused by the stellar pulsations as shown in Fig.~\ref{fig:M4_pulsation}. The pulsations generate streak features throughout the course of observations, entangling with the Rossiter-McLaughlin and planetary signal during transit. We empirically mitigated the stellar pulsation pattern in CCFs following \citet{Wyttenbach2020}. 
To achieve this, we suppose the pulsation features in the two-dimensional diagram (Fig.~\ref{fig:M4_pulsation}) stay static in terms of radial velocity, which can be approximated with positive or negative Gaussian profiles. We co-added the out-of-transit residual CCFs before ingress and after egress respectively, where the pulsation pattern appears symmetric before and after transit. We fit a Gaussian profile to the strongest peak in the combined out-of-transit residual CCFs and subtracted the fitted Gaussian component from all the individual out-of-transit CCFs, while the rest in-transit spectra remain untouched. Then the steps above were repeated to iteratively remove one Gaussian component at a time, until the major pulsation features were cleaned (5 iterations in our case, and the results are not sensitive to the number of iterations). 
The pulsation signal is mitigated while some structure remains visible in Fig.~\ref{fig:M4_pulsation} bottom panel, as we will also notice in Fig.~\ref{fig:ccfs}.

\subsection{Rossiter-McLaughlin reloaded} \label{sec:RM-reloaded}

\begin{figure}
   \centering
   \includegraphics[width=\hsize]{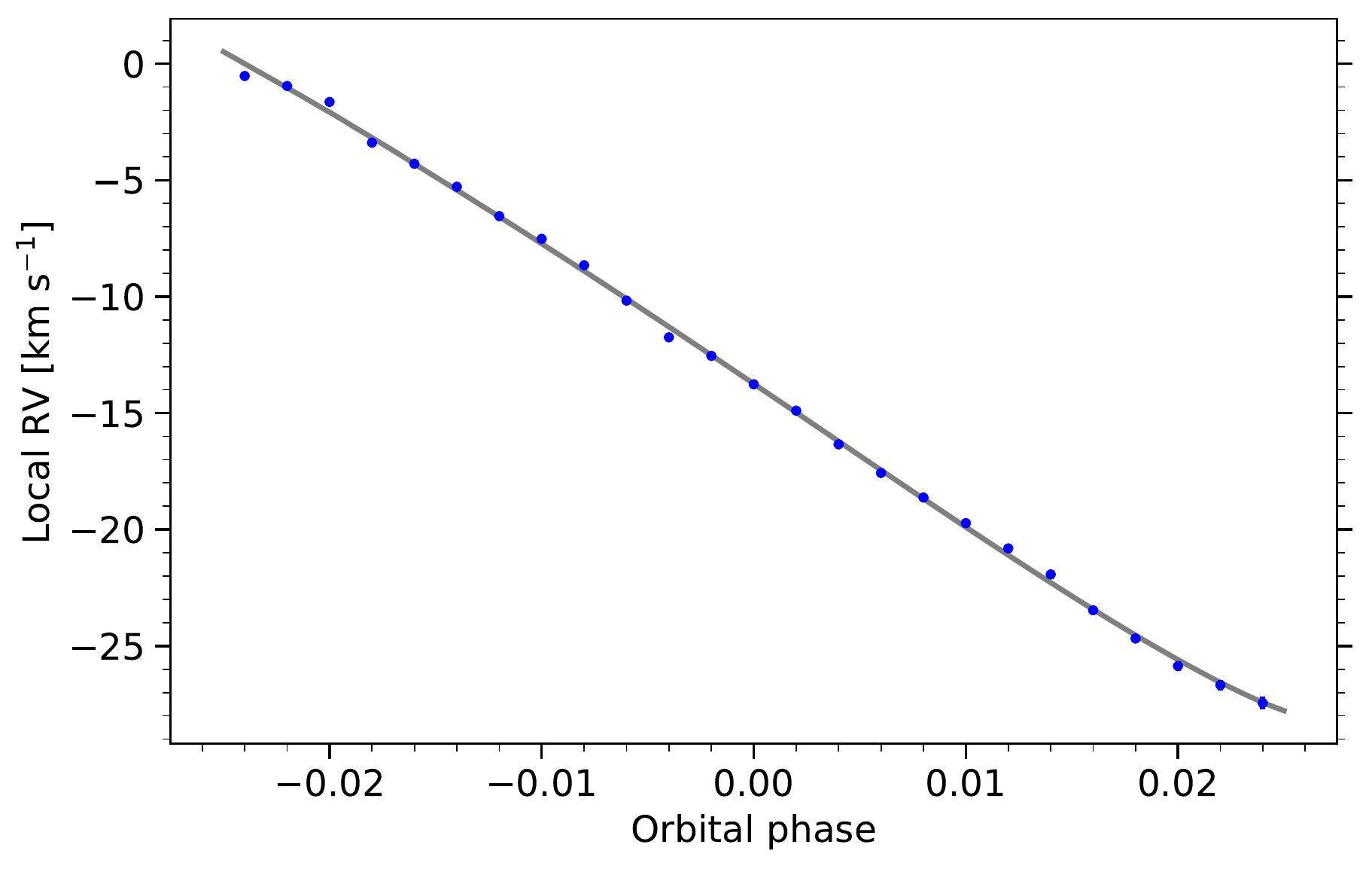}
      \caption{Local radial velocities (RVs) of the occulted stellar surface regions as measured from the yellow shadow in Fig.~\ref{fig:M4_pulsation}. The gray line is the best-fit RM reloaded model to the local RV data points.
              }
         \label{fig:RM-reloaded}
   \end{figure}

The Rossiter-McLaughlin (RM) effect (also known as the Doppler shadow) is the deformation of the stellar lines as a result of the planet blocking part of the stellar disk during transit. It encodes information of the stellar rotation \vsini and the projected misalignment angle $\lambda$ between the planet’s orbital axis and the star's rotation axis. 
We used the reloaded RM method \citep{Cegla2016} to model the doppler shift of the CCF profiles due to the occultation by the planet during transit. To extract RM information from the data, we combined the residual CCFs from both transits by binning in orbital phase with a step size of 0.002 and then fit the residual CCF at each phase with a Gaussian profile to determine the local RV of the occulted stellar surface. The measured local RVs plotted against orbital phases are shown in Fig.~\ref{fig:RM-reloaded}. 
We model the local RVs by computing the brightness-weighted average rotational velocity of the stellar surface blocked by the planet at each phase. Here we fixed the parameters such as $a/R_\ast$, $R_\mathrm{p}/R_\ast$ and $i_\mathrm{p}$ to the values listed in Table~\ref{tab:M4}, while making the spin-orbit angle $\lambda$, the stellar spin velocity $v$, inclination $i_\ast$, and differential rotation rate $\alpha$ free parameters. Fitting the model to the measured local RVs, we derived $\alpha=0.09\pm0.03$, $\lambda=250.34\pm0.14^{\circ}$, and \vsini= $46.2^{+7.7}_{-2.5}$ \kms. The values are consistent with the previous measurement of $\lambda=244.9_{-3.6}^{+2.7}$ and \vsini= $45.66_{-0.9}^{+1.1}$ \kms\ by \citet{Dorval2020}, with the slight difference in the spin-orbital angle likely resulting from the systematic differences in $P$ and $T_0$. Since we updated $P$ and $T_0$ from the simultaneous photometry as the spectroscopic observations, the updated epoch is more reliable for our analysis of the local RVs.
We caution that the uncertainties quoted here are underestimations because they did not account for systematics in the transit epoch and system parameters.

\subsection{Modeling RM and CLV effects} \label{sec:RM+CLV}

To disentangle the stellar signal from the planetary signal, we modeled the transit effects on stellar lines, including the Rossiter-McLaughlin (RM) and center-to-limb variation (CLV) effects, following \citep{Yan2017, Casasayas-Barris2019}.
We first computed synthetic stellar spectrum at different limb-darkening angles ($\mu$) using \texttt{Spectroscopy Made Easy} tool \citep{Valenti1996} with \texttt{VALD} line lists \citep{Ryabchikova2015}. The stellar disk was divided into cells of size 0.01 $R_\ast$ $\times$ 0.01 $R_\ast$, each assigned with a spectrum obtained by the interpolation to its corresponding $\mu$ and applying a radial velocity shift according to its local rotational velocity. We then integrated the whole stellar disk while excluding the region blocked by the planet during transit. We divided the integrated spectrum at each phase through the out-of-transit stellar spectrum, resulting in the model of RM+CLV effects such as shown in Fig~\ref{fig:line2d} (panel b). The system and stellar parameters used in the modeling are presented in Table~\ref{tab:M4}, including the best-fit parameters $\alpha=0.09$, $\lambda=250.3^{\circ}$, and \vsini= 48.6 \kms\ obtained via RM reloaded method in Section~\ref{sec:RM-reloaded}.

\subsection{Transmission spectrum} \label{sec:trasmission}
Using the telluric corrected 1D spectra, we extracted the transmission spectrum following the similar procedure in previous studies \citep[such as ][]{Wyttenbach2015, Casasayas-Barris2019}. It is summarised as follows. The spectra were median-normalised and shifted to the stellar rest frame. The out-of-transit spectra were co-added to build the master stellar spectrum, which was then removed from each individual spectrum via division. In the residuals, there remained sinusoidal wiggles as also seen in other ESPRESSO data \citep{Tabernero2021, Borsa2021, Casasayas-Barris2021, Kesseli2022}. We applied a Gaussian smoothing filter with a width of 5 \AA\ to each exposure and removed it to correct for the low-frequency noise.
Moreover, outliers exceeding 4$\sigma$ threshold in a sliding 25 \AA\ window were corrected through linear interpolation over nearby pixels.
Finally, we combined the data of both transits by binning in orbital phase with a step size of 0.002.

\begin{figure}
   \centering
   \includegraphics[width=\hsize]{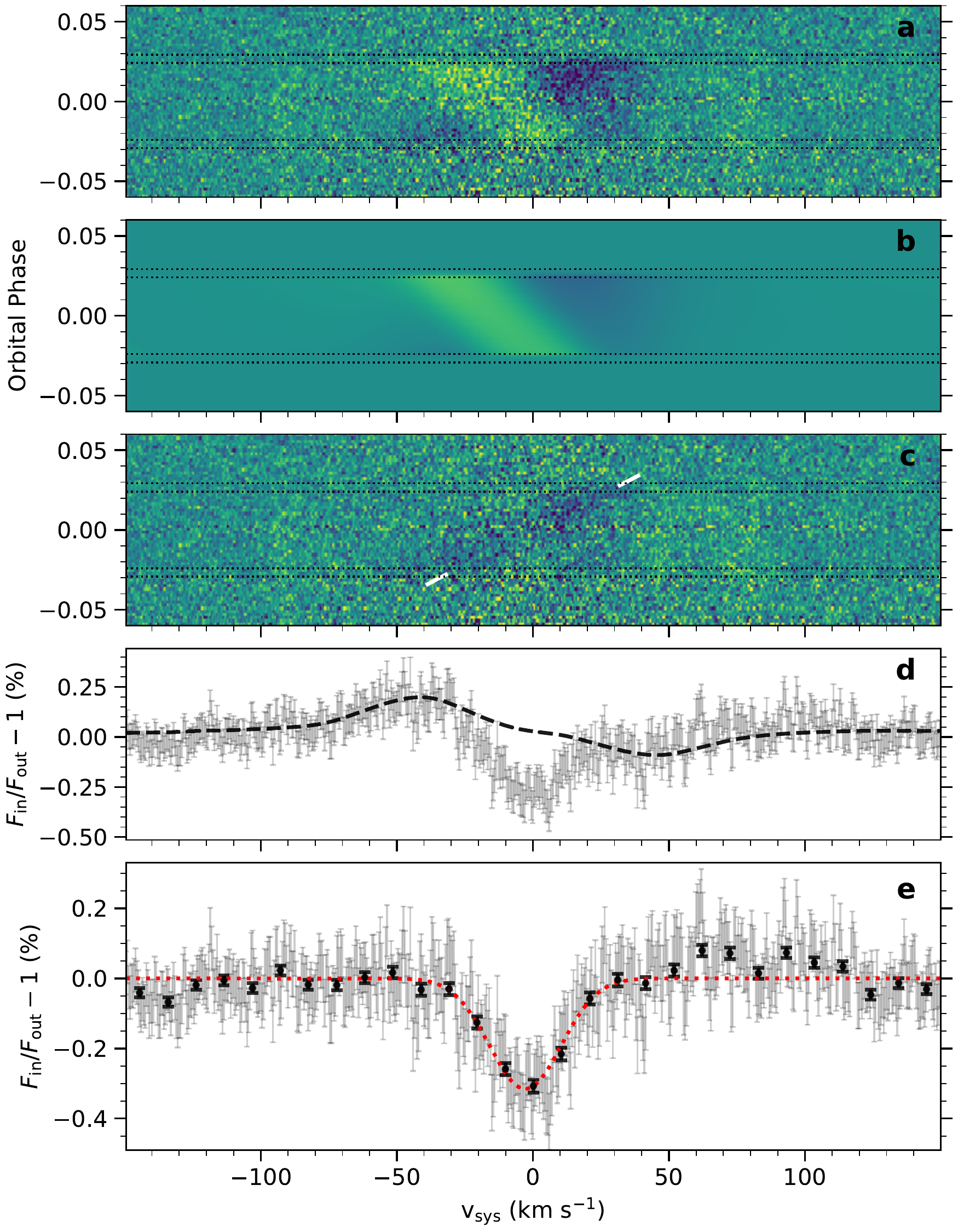}
      \caption{H$\alpha$ line transmission spectrum analysis of \M. \textit{Panel a}: residuals after removing the master out-of-transit spectrum and binning in orbital phase by 0.002, containing the stellar RM+CLV effects and the planetary absorption. \textit{Panel b}: best-fit RM+CLV model to the data in panel a. \textit{Panel c}: difference between the data (in panel a) and the RM+CLV model (in panel b) to isolate the planetary absorption signal, as traced by the slanted white line. \textit{Panel d}: transmission spectrum in grey obtained by average-combining the in-transit residuals in panel a (without the correction of RM+CLV effects) in the planet rest frame. The black dashed line is the RM+CLV model in panel b combined in the planet rest frame.  \textit{Panel e}: transmission spectrum in grey obtained by average-combining the cleaned residuals in panel c in the planet rest frame. The black points show the transmission spectrum binned every 15 data points. The red dotted line shows the best-fit Gaussian profile to the planetary absorption feature of H$\alpha$.
              }
         \label{fig:line2d}
   \end{figure}

The in-transit residuals at this stage contain the variation due to RM+CLV effects and the absorption of the planet. Following \citet{Yan2018}, we fit the data with a model composed of both the stellar effects (as detailed in Section~\ref{sec:RM+CLV}) and the planetary absorption signal (modeled as a Gaussian profile) assuming the expected planetary orbital motion amplitude ($K_{\rm p}$) as listed in Table~\ref{tab:M4}. The free parameters include the Gaussian amplitude ($h$), Gaussian width (FWHM), wind speed ($v_{\rm wind}$), and a scaling factor of the stellar effects ($f$) to account for the fact that the effective absorption radius can be larger than the nominal planet radius used in the RM+CLV model. The fitting process was performed with \texttt{PyMultiNest} \citep{Buchner2014}, a \texttt{Python} interface for the Bayesian inference technique \texttt{MultiNest} \citep{Feroz2009}. Once obtaining the best-fit values, we removed the stellar RM+CLV effects from the residuals, which were then average-combined in the planet rest frame to form the 1D transmission spectrum such as presented in Fig~\ref{fig:line2d} panel e.

\subsection{Cross-correlation analysis}

In addition to inspecting individual lines, we carried out cross-correlation analyses \citep{Snellen2010, Brogi2012} to co-add hundreds of absorption lines in the full range of the optical transmission spectrum to search for atoms \citep{Hoeijmakers2018, Hoeijmakers2019, Kesseli2022}.
We computed transmission models of different atoms and ions (\ion{Fe}{i}, \ion{Fe}{ii}, \ion{Mg}{i}, etc.) for cross-correlation analysis using the radiative transfer tool \texttt{petitRADTRANS} \citep{Molliere2019}.
Here we assumed an isothermal temperature profile of 2500 K, a continuum level of 1 mbar, and a gray cloud deck at 1 mbar. The volume mixing ratios (VMR) were set to the solar abundance. 
We utilized the formula for cross-correlation as presented in \citep{Hoeijmakers2018, Hoeijmakers2019, Hoeijmakers2020b},
\begin{equation} \label{eq:ccf}
    c (v, t) = \frac{\sum_{i} x_i(t) T_i(v)}{\sum_{i} T_i(v)},
\end{equation}
where $x_i(t)$ is the observation at time $t$. $T_i(v)$ is the transmission model of the species shifted to a radial velocity $v$, such that the CCF effectively is a weighted average of multiple absorption lines, representing the average strength of the absorption.
Following this convention allows us to compare the line strengths across different ultra-hot Jupiters presented in previous cross-correlation studies. However, we caution that such CCF amplitudes depend on the specific models used. This will be further discussed in the comparison to other planets in Section~\ref{sec:correlation}.

The transmission templates were cross-correlated with the telluric corrected spectra in the wavelength range of 380-685 nm (beyond which the spectra are heavily contaminated by telluric lines, therefore excluded in the analysis). This led to the stellar CCFs dominated by signals from the stellar spectra. We then obtained the residual CCFs by removing the average out-of-transit CCF and mitigated the stellar pulsation pattern following Section~\ref{sec:pulsation}. Similar as the transmission spectrum, the residual CCFs contains both the stellar RM+CLV effects and the planetary signal. We carried out the same cross-correlation analysis on the synthetic stellar spectra computed in Section~\ref{sec:RM+CLV} to simulate the RM+CLV contribution to the residual CCFs, which was multiplied by a scaling factor $f$ and then removed from the data to obtain the final CCFs originating from the planetary absorption. The values of free parameters including $f$, the Gaussian amplitude of the absorption signal $h$, the Gaussian FWHM and the central velocity offset $v_{\rm wind}$ were determined similarly as presented in Section~\ref{sec:trasmission}.

\section{Results} \label{sec:result}

\subsection{Detection of individual lines of \texorpdfstring{ \ion{H}{i}, \ion{Na}{i}, \ion{Ca}{ii}, \ion{Mg}{i}, \ion{Fe}{i}, \ion{Fe}{ii}}{}} \label{sec:result_lines}

\begin{figure*}
   \centering
   \includegraphics[width=\hsize]{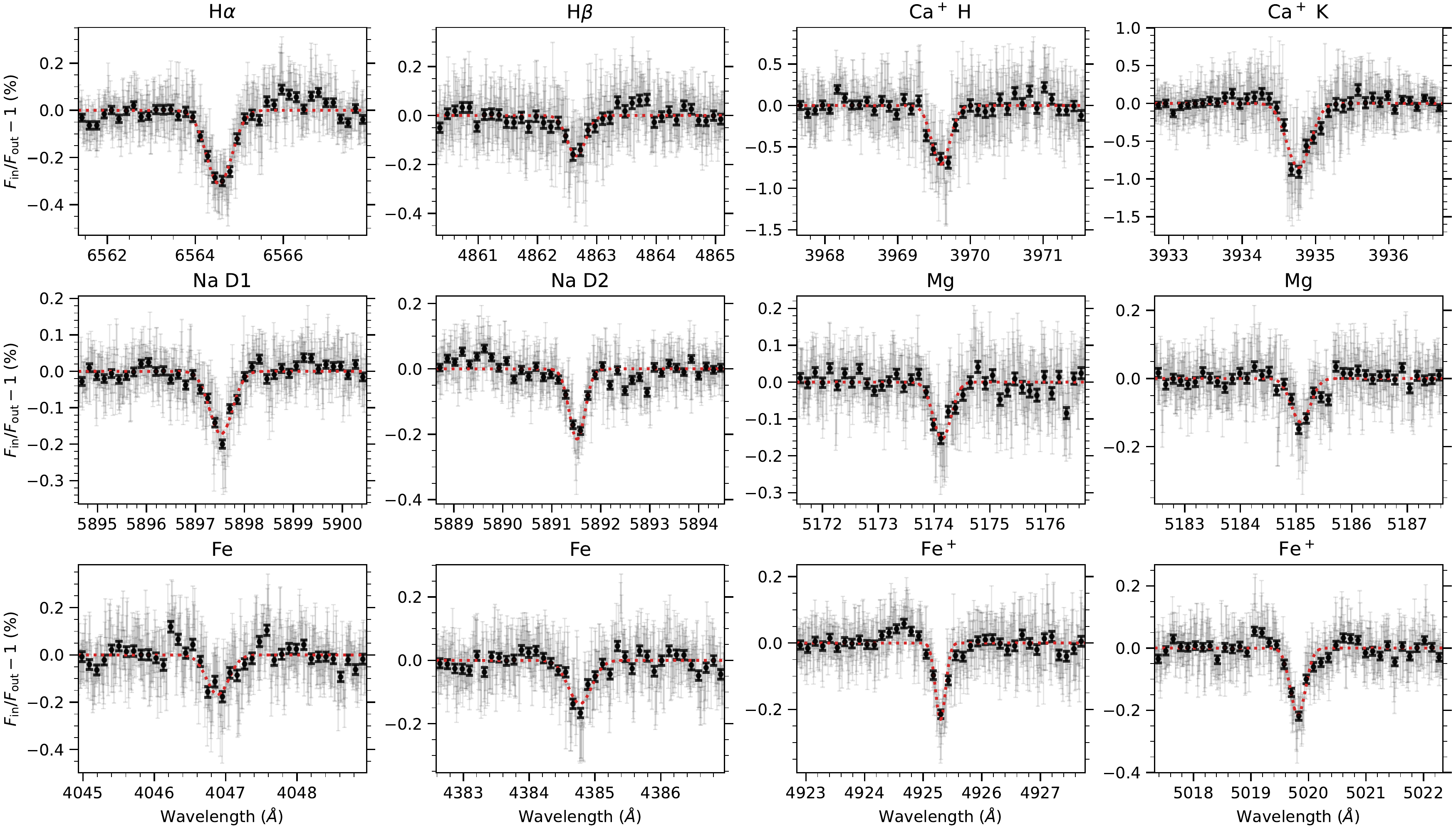}
      \caption{Transmission spectrum around H$\alpha$, H$\beta$, \ion{Na}{i} D1\&D2, \ion{Ca}{ii} H\&K, \ion{Mg}{i}, \ion{Fe}{i}, and \ion{Fe}{ii} lines, averaged over both nights of observation. The black points show the transmission spectrum binned every 15 grey data points. The red dotted line shows the best-fit Gaussian profile to the planetary absorption feature.
              }
         \label{fig:line1d}
   \end{figure*}

We report the detection of individual absorption lines of H$\alpha$, H$\beta$, \ion{Na}{i} D1\&D2, \ion{Ca}{ii} H\&K, \ion{Mg}{i}, \ion{Fe}{i}, and \ion{Fe}{ii} in \M. The transmission spectra around these absorption lines are shown in Fig.~\ref{fig:line1d}, and the measured properties are summarised in Table~\ref{tab:lines}.

The centre of the absorption features generally agree with zero, while H$\alpha$, H$\beta$, and \ion{Na}{i} doublet appear blueshifted by up to $\sim$-4 \kms, which is usually interpreted as the evidence of day-to-night side wind \citep{Snellen2010, Brogi2016, Hoeijmakers2018, Casasayas-Barris2019, Seidel2021}. The various velocity offsets of different species may indicate that the lines probe distinct atmospheric layers dominated by different dynamic processes.

The wind velocity offset for the \ion{Na}{i} doublet differs from each other by $\sim 2\sigma$. We suppose this may be systematics as a result of multiple sodium absorptions by the interstellar medium located around 13 to 24 \kms away from the line center. We mitigated the effect by excluding this velocity range at the barycentric rest-frame when calculating the transmission spectrum, but some artefacts might still persist to contribute to the line offset. 

We also note in Fig.~\ref{fig:Ha} panel c the 'gap' in the absorption signal when the planetary trail intersects the Doppler shadow, meaning that the planetary transmission lines overlap with the stellar lines from the region blocked by the planet. At this moment, the effective size of the planet appears larger because of the absorption, therefore enhancing the RM effect. This is not accounted for in our RM+CLV modelling, so we commonly see such under-correction that leads to the gap near the overlapping orbital phases. 
We quantified the effect of the under-correction on the planetary absorption depths by excluding the overlapping orbital phases (e.g. from -0.015 to 0.015) when fitting the planetary signal and co-adding the transmission spectra. We found that the absorption amplitude increases by $\sim$20\%-25\% for lines such as H$\alpha$, sodium, and ionised iron, typically around 2$\sigma$ of our measurements. The nominal uncertainties shown in Table~\ref{tab:lines} did not account for such systematic noise, therefore likely to be underestimations.

\begin{table*}
\caption{Summary of individual line detection in the transmission spectrum of \M.}
\label{tab:lines}      
\centering                         
\begin{tabular}{c c c c c c c c}     
\hline\hline                
                    
Line & $\lambda_0$ (\AA) & $h$ (\%) &  $N_\mathrm{sc}$  & S/N  & $f$ & FWHM (\kms) & $v_{\rm wind}$ (\kms)\\    
\hline  
H$\alpha$ & 6564.61 & $ -0.317 \pm 0.021 $  & $ 48.4 \pm 3.3 $   & 14.8 & $ 1.89 $  & $ 31.4 \pm 2.4 $  & $ -3.0 \pm 1.0 $  \\
H$\beta$  & 4862.72 & $ -0.143 \pm 0.030 $  & $ 21.9 \pm 4.5 $   & 4.8 & $ 1.49 $  & $ 27.2 \pm 8.1 $  & $ -4.5 \pm 2.3 $  \\
\ion{Ca}{ii} H & 3969.59 & $ -0.705 \pm 0.082 $  & $ 107.6 \pm 12.6 $   & 8.6 & $ 1.48 $  & $ 23.0 \pm 3.0 $  & $ 0.2 \pm 1.3 $  \\
\ion{Ca}{ii} K  & 3934.77 & $ -0.844 \pm 0.082 $  & $ 128.9 \pm 12.5 $   & 10.3 & $ 1.31 $  & $ 29.6 \pm 3.1 $  & $ 0.4 \pm 1.5 $  \\
\ion{Na}{i} D1 & 5897.55 & $ -0.168 \pm 0.014 $  & $ 25.6 \pm 2.2 $   & 11.9 & $ 1.32 $  & $ 28.6 \pm 2.8 $  & $ -1.8 \pm 1.1 $  \\
\ion{Na}{i} D2 & 5891.58 & $ -0.214 \pm 0.017 $  & $ 32.7 \pm 2.6 $   & 12.4 & $ 1.37 $  & $ 19.9 \pm 2.1 $  & $ -3.6 \pm 0.7 $  \\
\ion{Mg}{i}  & 5174.12 & $ -0.151 \pm 0.017 $  & $ 23.0 \pm 2.6 $   & 8.7 & $ 1.04 $  & $ 22.6 \pm 3.5 $  & $ 1.5 \pm 1.3 $  \\
\ion{Mg}{i}  & 5185.05 & $ -0.125 \pm 0.019 $  & $ 19.1 \pm 2.9 $   & 6.5 & $ 1.08 $  & $ 25.2 \pm 5.8 $  & $ 2.3 \pm 1.6 $  \\
\ion{Fe}{i} & 4046.96 & $ -0.162 \pm 0.025 $  & $ 24.8 \pm 3.8 $   & 6.6 & $ 1.04 $  & $ 27.8 \pm 4.4 $  & $ -5.0 \pm 2.1 $  \\
\ion{Fe}{i} & 4384.78 & $ -0.135 \pm 0.016 $  & $ 20.6 \pm 2.5 $   & 8.3 & $ 1.16 $  & $ 29.2 \pm 4.1 $  & $ -0.0 \pm 1.6 $  \\
\ion{Fe}{ii} & 4925.30 & $ -0.226 \pm 0.022 $  & $ 34.5 \pm 3.3 $   & 10.3 & $ 1.13 $  & $ 13.1 \pm 1.7 $  & $ -0.1 \pm 0.6 $  \\
\ion{Fe}{ii} & 5019.83 & $ -0.211 \pm 0.020 $  & $ 32.2 \pm 3.0 $   & 10.8 & $ 1.11 $  & $ 19.1 \pm 2.5 $  & $ -0.7 \pm 0.8 $  \\
\hline                                  
\end{tabular}
\tablefoot{$\lambda_0$ is the central wavelength of the line in vacuum. $h$, FWHM, and $v_{\rm wind}$ are the amplitude, width, and center of the best-fit Gaussian profile to the planetary absorption. S/N is simply calculated as the value of $h$ divided by its uncertainty. $N_\mathrm{sc}$ represents the number of atmospheric scale heights that the peak absorption corresponds to. $f$ is the scaling factor applied to the stellar RM+CLV model.}
\end{table*}

\subsection{Detection of species in cross-correlation}
\begin{figure*}
   \centering
   \includegraphics[width=\hsize]{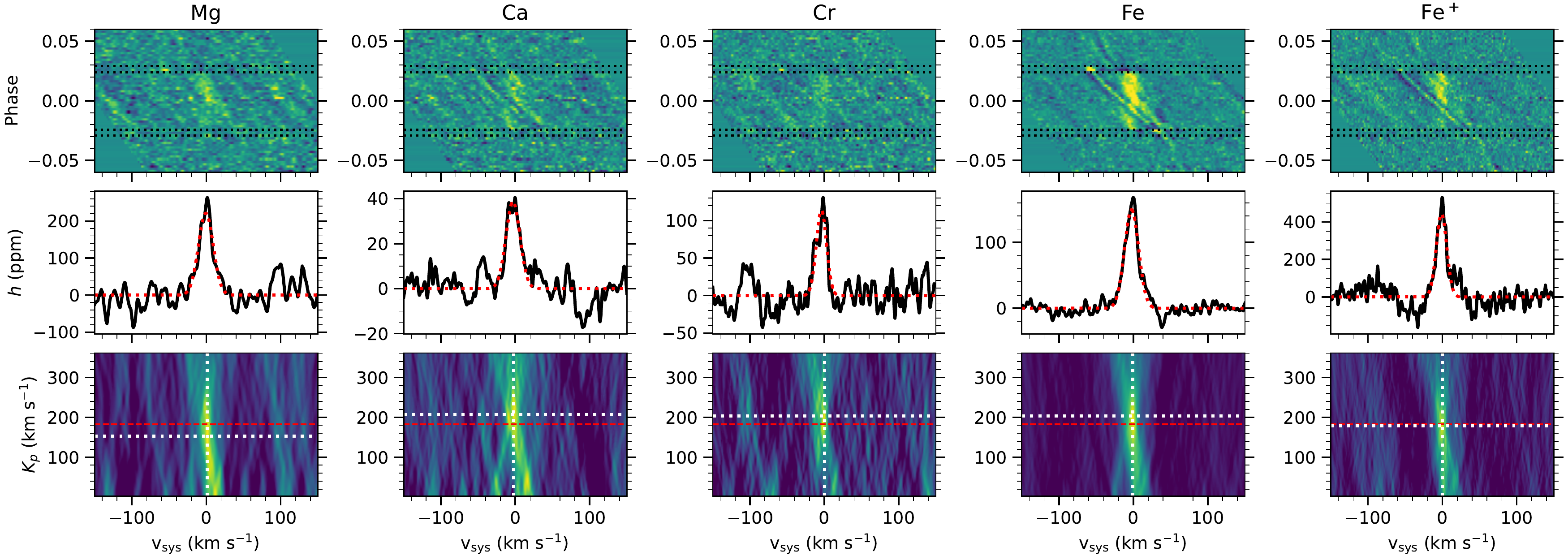}
      \caption{ Cross-correlation signatures of \ion{Mg}{i}, \ion{Ca}{i}, \ion{Cr}{i}, \ion{Fe}{i}, and \ion{Fe}{ii}, averaged over both nights of observation. 
      \textit{Top row}: two-dimensional residual cross-correlation functions at the planet rest frame corresponding to the expected orbital velocity $K_{\rm p}$. The black dotted lines denote the timings of four transit contacts. The vertical signature in yellow around zero velocity illustrates the planetary transmission signal. The narrow slanted shape is caused by imperfect correction of the stellar RM effect.
      \textit{Middle row}: one-dimensional CCF by co-adding all in-transit residual CCFs shown in the top panel.
      The red dotted line shows the best-fit Gaussian profile to the planetary signal, the parameters of which are summarised in Table~\ref{tab:ccfs}.
      \textit{Bottom row}: a stack of co-added CCFs assuming a range of planetary velocity $K_{\rm p}$ (in y axis). The white dotted lines indicate the maximum in the $K_{\rm p}$-$V_\mathrm{sys}$ map, and the red dashed line marks the expected $K_{\rm p}$.
              }
         \label{fig:ccfs}
   \end{figure*}

In addition to elements with individual lines detected in the transmission spectrum, we performed cross-correlation analyses for a range of atoms, ions, and molecules, guided by their observability at high spectral resolution as presented in \citet{Kesseli2022}. Here we present the detection of \ion{Mg}{i}, \ion{Ca}{i}, \ion{Cr}{i}, \ion{Fe}{i}, and \ion{Fe}{ii} in Fig.~\ref{fig:ccfs} and Table~\ref{tab:ccfs}. We found no evidence of other species such as \ion{Ti}{i}, \ion{Ti}{ii}, \ion{V}{i}, \ion{V}{ii}, \ion{Mn}{i}, \ion{Co}{i}, \ion{Ni}{i}, TiO, VO.

The lack of detection of \ion{Ti}{i}, \ion{Ti}{ii}, and TiO are commonly seen in UHJs, although not well understood. 
Temperatures in the atmosphere seem to play a key role in determining the chemical composition. For example, KELT-9~b shows strong \ion{Ti}{ii} and no \ion{Ti}{i}. Therefore the lack of \ion{Ti}{i} is likely attributed to the dominant ionization at the extreme temperature of 4000 K \citep{Hoeijmakers2018}. 
The detection of \ion{Ti}{i}, \ion{Ti}{ii}, and TiO was found in WASP-189b \citep{Prinoth2022} with a temperature of $\sim$2700 K. For other UHJs with slightly lower temperatures (including WASP-76~b, WASP-121~b, HAT-P-70~b MASCARA-2~b, and \M), there is no conclusive detection of Ti in any form. This has been proposed to be due to Ti being trapped in condensates on the night side of those cooler planets \citep{Spiegel2009, Parmentier2013, Kesseli2022}.

\begin{table*}
\caption{Summary of cross-correlation detection.}
\label{tab:ccfs}      
\centering                         
\begin{tabular}{c c c c c c c}     
\hline\hline                
Species & S/N & $K_{\rm p}$ (\kms) & $h$ (\%) & $f$ & FWHM (\kms) & $v_{\rm wind}$ (\kms)  \\    
\hline                      

\ion{Mg}{i} & 6.3  & $ 153 ^{+60}_{-62} $  & $ 0.0221 \pm 0.0009 $  & 0.54   & $ 23.8 \pm 0.8 $ & $ -0.5 \pm 1.6 $  \\
\ion{Ca}{i} & 6.3  & $ 207 ^{+53}_{-60} $  & $ 0.0039 \pm 0.0001 $  & 0.69  & $ 22.7 \pm 1.2 $ & $ -3.2 \pm 1.5 $  \\
\ion{Cr}{i}  & 6.7  & $ 204 ^{+14}_{-45} $  & $ 0.0114 \pm 0.0006 $  & 0.42 & $ 16.3 \pm 0.6 $  & $ -3.9 \pm 1.0 $  \\
\ion{Fe}{i} & 25.3  & $ 204 ^{+13}_{-21} $  & $ 0.0150 \pm 0.0001 $  & 0.35 & $ 22.5 \pm 0.2 $  & $ -2.4 \pm 0.4 $  \\
\ion{Fe}{ii}   & 11.8  & $ 179 ^{+14}_{-12} $  & $ 0.0444 \pm 0.0011 $  & 0.62  & $ 16.5 \pm 0.4 $  & $ -1.2 \pm 0.6 $  \\
\hline                                  
\end{tabular}
\tablefoot{S/N is the signal-to-noise ratio at the maximum in the $K_{\rm p}$-$V_\mathrm{sys}$ map as shown in Fig.~\ref{fig:ccfs} bottom row. The noise level is measured in the map as the standard deviation in the velocity range of (-150, -75) and (75, 150) \kms, away from the peak signal. The uncertainties of parameters are computed following the method as described in \citet{Kesseli2022}. $h$, $f$, FWHM, and $v_{\rm wind}$ are defined the same as in Table~\ref{tab:lines}.}
\end{table*}  

\subsection{Neutral and Ionized iron in MASCARA-4b and other UHJs}

Here we focus on properties of \ion{Fe}{i} and \ion{Fe}{ii} absorption in \M and draw comparisons with previous detection in other ultra-hot Jupiters.

The absorption strength of \ion{Fe}{ii} exceeds \ion{Fe}{i}, and is more than an order of magnitude higher than what is predicted under the assumption of a hydrostatic atmospheric model.
This has been commonly seen in other UHJs such as KELT-9b \citep{Hoeijmakers2019}, MASCARA-2b \citep{Hoeijmakers2020b}, HAT-P-70b \citep{Bello-Arufe2022} and WASP-189b \citep{Prinoth2022}.
The contribution from the hydrostatic region of atmospheres is often too small to account for the measured absorption level of \ion{Fe}{ii}.
Several deviations from the model assumption may help explain the strong absorption by \ion{Fe}{ii}, for instance, photochemistry in the upper atmosphere, non local thermodynamic equilibrium (non-LTE) effects, hydrodynamic outflows \citep{Huang2017, Hoeijmakers2019, Prinoth2022}. These all suggest that \ion{Fe}{ii} traces the upper atmospheres, higher than \ion{Fe}{i} does. In particular, hydrodynamic outflows can raise the species to the extended upper atmosphere, followed by progressive ionisation of \ion{Fe}{i} in the exosphere, giving rise to strong \ion{Fe}{ii} absorption features. Moreover, the strong lines of \ion{Fe}{ii} and \ion{Mg}{ii} observed in the near-ultraviolet (NUV) in WASP-121b also provide evidence of the exospheric origins of the ionic species \citep{Sing2019}.

As shown in Table~\ref{tab:ccfs}, the FWHM of \ion{Fe}{i} signal in \M is larger than that of \ion{Fe}{ii}. This has also been observed in MASCARA-2b \citep{Hoeijmakers2020b}, while HAT-P-70b shows the opposite trend that \ion{Fe}{ii} is broader than \ion{Fe}{i} \citep{Bello-Arufe2022}. This discrepancy of line width adds to the indication that neutral and ionised iron may probe different region in the atmosphere. \ion{Fe}{i} traces deep down layers, while \ion{Fe}{ii} originates from the upper atmosphere, where the distinct dynamic regimes contribute to the different line widths and radial velocities \citep{Showman2013, Louden2015, Brogi2016, Seidel2019, Hoeijmakers2019, Bello-Arufe2022}. For instance, super-rotational jets may be present in the lower atmospheres of \M and MASCARA-2b, resulting in broadened \ion{Fe}{i} signatures. Whereas, the atmosphere of HAT-P-70b may undergo strong outflow in the upper atmosphere, broadening \ion{Fe}{ii} signal instead.

\section{Discussion} \label{sec:discussion}

   \begin{figure*}
   \centering
   \includegraphics[width=\hsize]{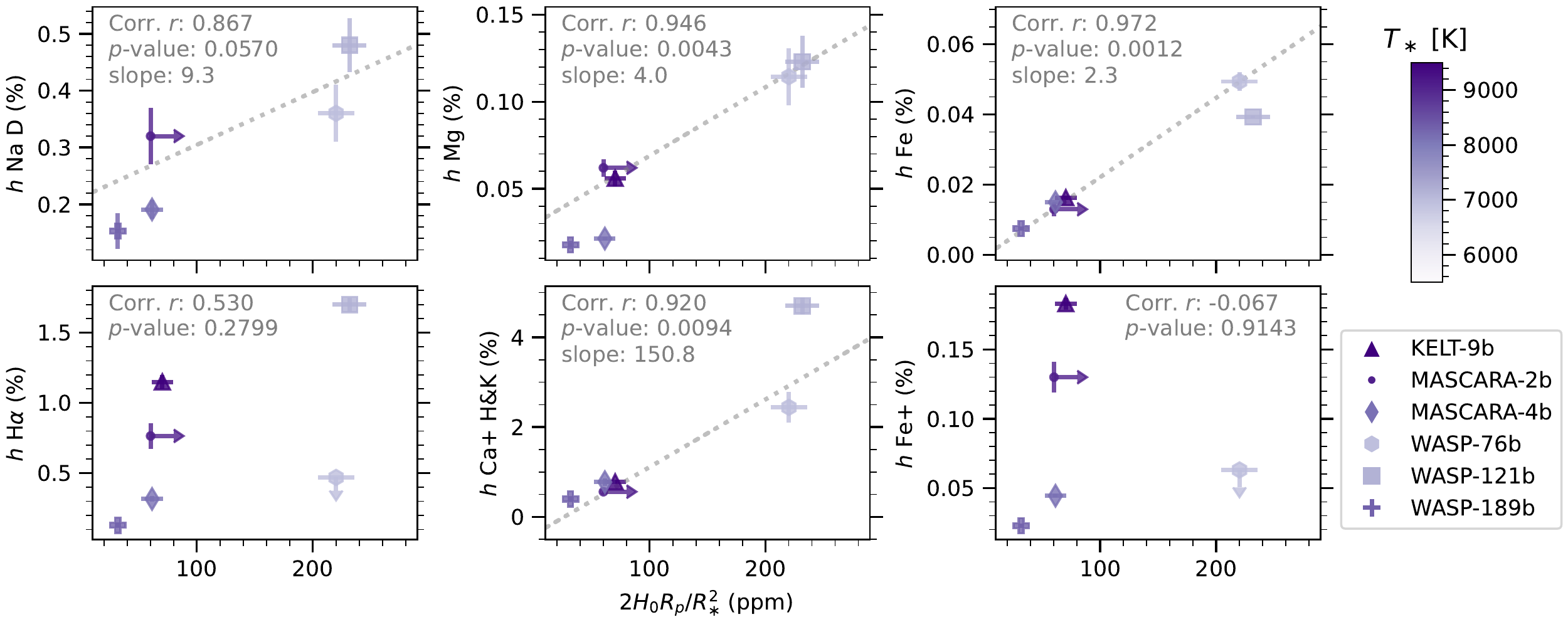}
      \caption{Correlation of observed line strengths ($h$) of different species with expected transmission strengths of absorbers extending one atmospheric scale height ($2H_0R_p/R_\ast^2$) for different UHJs. The color saturation of data points represents the temperature of the host star. The data used in this plot are compiled in Table~\ref{tab:UHJs}. Planets with high upper limits on mass (e.g. HAT-P-70b) are omitted in the plot. The Pearson correlation coefficients $r$ and the $p$-values for testing non-correlation are calculated for each species. For metals such as \ion{Mg}{i} and \ion{Fe}{i}, $r$ close to 1 and $p$-value close to 0 indicate strong linear correlation of absorption amplitudes with scale heights. The linear fit to the data points is delineated with dotted gray lines, the slope of which represents the number of scale heights that a certain species probes.
              }
         \label{fig:scaleheight}
   \end{figure*}
   
\subsection{Disentangling the hydrostatic atmosphere and extended exosphere of UHJs} \label{sec:correlation}

As the detection of atoms with high-resolution transmission spectroscopy accumulates quickly, a small sample of ultra-hot Jupiters starts to build up, providing us with the opportunity to study potential trends of atomic signatures in the UHJ population.
Among various species, \ion{H}{i}, \ion{Na}{i}, \ion{Mg}{i}, \ion{Ca}{ii}, \ion{Fe}{i}, and \ion{Fe}{ii}, have been commonly detected in a handful of UHJs, including KELT-9b \citep{Yan2018, Cauley2019, Borsa2019, Hoeijmakers2019, Yan2019, Turner2020, Wyttenbach2020}, MASCARA-2b \citep{Casasayas-Barris2020, Hoeijmakers2020b, Nugroho2020, Stangret2020}, WASP-121b \citep{Cabot2020, Gibson2020, Hoeijmakers2020a, Borsa2021, Merritt2021}, WASP-76b \citep{Tabernero2021, Seidel2021, Casasayas-Barris2021, Kesseli2022}, WASP-189b \citep{Prinoth2022}, HAT-P-70b \citep{Bello-Arufe2022}. We compile properties of the detections (including the transmission amplitude $h$ and the FWHM of each species) in Table~\ref{tab:UHJs}.

Fig.~\ref{fig:scaleheight} shows the sample of UHJs with the absorption amplitude ($h$) of each species plotted against the typical transmission strength of absorbers extending one scale height ($2H_0R_p/R_\ast^2$). Under the assumption of hydrostatic atmospheres, we expect the line strength of one species to be proportional to the typical transmission amplitude $2H_0R_p/R_\ast^2$, if the absorption forms at a similar pressure level. In this case, the slope of the linear correlation represents the vertical extent of the atom in UHJ atmospheres in units of scale height $H_0$. According to Fig.~\ref{fig:scaleheight}, neutral metal species such as \ion{Mg}{i}, \ion{Fe}{i}, (and possibly \ion{Na}{i}) follow the trend well, with the Pearson correlation coefficients $r$ close to 1. The number of scale heights (see the slopes in Fig.~\ref{fig:scaleheight}) probed by \ion{Na}{i}, \ion{Mg}{i}, and \ion{Fe}{i} decreases with the atomic mass of the element. Another underlying assumption for the linear correlation is that the abundances of the neutral species do not vary significantly in all the UHJs. The good correlations shown in Fig.~\ref{fig:scaleheight} seem to hint at the validity of this hypothesis, which could be verified by future retrieval analysis to constrain the abundances in these UHJs.

On the other hand, H$\alpha$ and \ion{Fe}{ii} are two apparent exceptions of the correlation. In particular, although the scale height of WASP-76b is large, only upper limits have been estimated for H$\alpha$ and \ion{Fe}{ii} \citep{Casasayas-Barris2021, Kesseli2022}. Instead, this agrees with the argument that the absorption of H$\alpha$ probes extended upper atmospheres where the hydrostatic and LTE assumptions no longer apply. 
Hence the \ion{Fe}{ii} signatures, with such a similar behaviour as H$\alpha$, likely also originate from UHJ exospheres.

\begin{figure}
  \centering
  \includegraphics[width=\hsize]{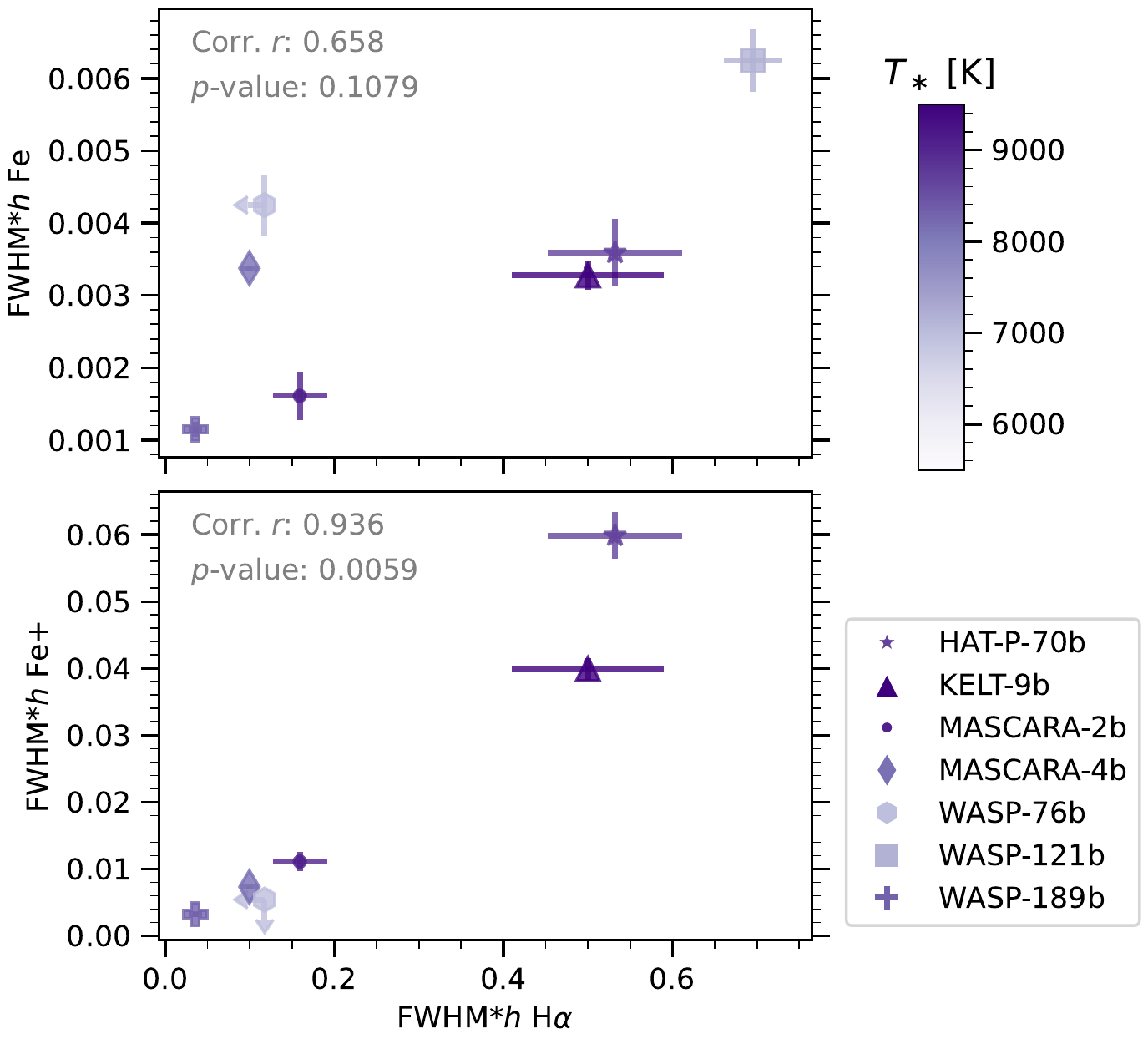}
      \caption{Absorption level of \ion{Fe}{i} or \ion{Fe}{ii} against H$\alpha$ in UHJs. The absorption level, taking into account both the amplitude and width of the atomic absorption profile (as a proxy for the equivalent width), is proportional to the total number of the absorbing species in the optically thin limit. We note the strong correlation of H$\alpha$ with \ion{Fe}{ii} (bottom panel), but less significant with \ion{Fe}{i} (top panel). 
              }
         \label{fig:Ha}
  \end{figure}

In this light, we find a positive correlation of absorption signals between \ion{Fe}{ii} and H$\alpha$ as shown in Fig.~\ref{fig:Ha}, consolidating that they probe the similar atmospheric region and process.
The planet WASP-76b, with only an upper limit detection of the H$\alpha$ absorption, also show no evidence of \ion{Fe}{ii}, well in line with the correlation. WASP-121b is a baffling case where the detection of \ion{Fe}{ii} in the optical is debated (the tentative detection was claimed in \citet{Ben-Yami2020, Borsa2021, Merritt2021}, but contradicted in \citet{Hoeijmakers2020a, Gibson2020}), while the strong detection in the NUV \citep{Sing2019} does suggest its presence in the extended exosphere up to $2R_p$. Based on the strong H$\alpha$ detection in WASP-121b \citep{Cabot2020, Borsa2021}, we expect significant \ion{Fe}{ii} absorption if it follows the trend. Further observations are needed to unravel this.
We also note that in order to ensure the trend is not obstructed by the model-dependency of the CCF signal, we compared individual \ion{Fe}{ii} lines in transmission spectra of KELT-9b \citep{Cauley2019, Hoeijmakers2019}, MASCARA-2b \citep{Casasayas-Barris2020}, and \M (this work), and confirm that the correlation with H$\alpha$ still holds.

Therefore, our comparison of atomic transmission features among the UHJ population hints at two distinct regimes of origin, the hydrostatic lower atmosphere and the extended exosphere. The linear correlation between the absorption strengths of metals (such as \ion{Na}{i}, \ion{Mg}{i}, and \ion{Fe}{i}), and the expected transmission amplitudes under the hydrostatic assumption validates their origin from the hydrostatic lower atmosphere. On the other hand, hydrogen and ions such as \ion{Fe}{ii} deviate from the scale height correlation, possibly because of the prevailing contribution from hydrodynamic outflows in the upper atmosphere. The positive relation of absorption strength between \ion{Fe}{ii} and H$\alpha$ is further indicative of their exospheric origins (as discussed in Section~\ref{sec:exosphere}). The overall picture for \ion{Ca}{ii} is less clear, probably involving contribution of both regimes. The absorption strengths of \ion{Ca}{ii} are commonly large enough to be attributed to the extended upper atmospheres, sometimes even beyond the Roche lobe \citep{Borsa2021, Bello-Arufe2022}. Yet we find no clear correlation between \ion{Ca}{ii} and H$\alpha$, while the linear trend with scale heights still hold to some extent (see Fig.~\ref{fig:scaleheight}). Our speculation is that both lower and upper atmospheric regimes contribute to the absorption, considering \ion{Ca}{i} atoms are prone to be readily ionized in the lower atmosphere in contrast to \ion{Fe}{i}, the ionization of which may only be significant in the upper atmosphere. A larger sample size is required to draw more solid conclusions.

We caution that studying an ensemble of UHJs is challenging because of two aspects.
First, it is difficult to account for the uncertainty of measurements either from different instruments or from different data reduction. For instance, systematic differences have been found in H$\alpha$ in KELT-9b \citep{Yan2018, Cauley2019, Turner2020, Wyttenbach2020} and MASCARA-2b \citep{Casasayas-Barris2019}. However, the systematics are not expected to be large enough to break the general trend that we show here.
Second, the model-dependency of cross-correlation signals presents challenges for the comparison of the UHJ population. Previous studies of individual UHJs use cross-correlation templates that are modeled differently. A standard set of models such as presented in \citet{Kitzmann2022} will be beneficial if it can be commonly used in such analyses. Yet the choice of temperature in the model affects the relative weights assigned to weak versus strong lines, which changes the amplitude of CCFs by up to a factor of two.

\subsection{Hydrodynamic exospheres as probed via \texorpdfstring{H$\alpha$}{} and ions}\label{sec:exosphere}

The positive correlation of absorption signals between \ion{Fe}{ii} and H$\alpha$ as shown in Fig.~\ref{fig:Ha} indicates that both species trace the similar atmospheric region in UHJs.
\ion{H}{i} and \ion{He}{i} absorption lines have been previously modeled as a probe for the escaping exosphere \citep{Huang2017, Allan2019, Garcia-Munoz2019, Wyttenbach2020, Yan2021, Oklopcic2018, Lampon2021, DosSantos2022}, which is expected to be the consequence of hydrodynamic outflows driven by stellar X-ray, extreme Ultraviolet (EUV) or NUV radiation. The driving mechanism may depend on the stellar type, as early A type stars are not expected to emit strongly at EUV \citep{Fossati2018}, while having high levels of NUV flux \citep{Garcia-Munoz2019}. 
The modeling of outflows also extends to heavy atoms such as C, O, Si by \citet{Koskinen2013}, suggesting that heavy elements dragged to the upper atmospheres stay well mixed as a result of collisions with rapidly escaping hydrogen. 
\citet{Gebek2020} modeled Na and K in evaporative exospheres to interpret high-resolution transmission observations. 
Furthermore, the strong \ion{Mg}{i}, \ion{Mg}{ii}, and \ion{Fe}{ii} lines in the NUV are discussed as tracers of upper atmospheres and hydrodynamic escapes \citep{Bourrier2014, Sing2019, Dwivedi2019}, while the optical lines have not been explored. 
Based on the presented correlation of absorption signals between \ion{Fe}{ii} and H$\alpha$, we suggest that \ion{Fe}{ii} lines in the optical also probe exospheres of UHJs. Therefore, further modeling of \ion{Fe}{ii} in exosphere can help constrain the structure of upper atmosphere, outflows, and the mass loss process.

Without the intention to model any particular atmosphere in detail, we aim to get insights into the trend in the UHJ population by some simplified estimations as follows.
We examine the role of hydrodynamic outflows and photoionisation of atoms in the absorption signals of \ion{H}{i} and \ion{Fe}{ii} in UHJs.
For a rough estimation, we assume the extended exosphere as a homogeneous optically thin cloud subject to stellar high energy radiation. 
In the optically thin limit, the absorption level is proportional to the total mass of the absorbing material regardless of the shape of the cloud \citep{Hoeijmakers2020a, Gebek2020}. Hence, the absorption signal depends on the mass loss rate $\dot{M}$ that determines the inflow of absorbing material to the exosphere and the ionisation degree $f_X$ of the element.
More details of the estimation can be found in Appendix~\ref{app:estimation}.
We find that the photoionisation plays a marginal role in the discrepancy of absorption strengths because the degree of ionization is expected to vary by less than a factor of a few among different UHJs. Instead, the amount of absorption is dominated by the planetary mass loss $\dot{M}$ driven by the stellar EUV or NUV flux, which usually varies by orders of magnitudes from system to system. Therefore, we argue that the dominant outflow drives the positive correlation between the H$\alpha$ and \ion{Fe}{ii} absorption (see Fig.~\ref{fig:Ha}), and they likely trace the exospheres of UHJs.
The absorption level reflects the properties of individual planets such as the mass loss rate and the irradiation environment. 
 
Although the sample size of UHJs with detailed spectral characterisation is still small, we suggest that the correlation between \ion{Fe}{ii} and H$\alpha$ absorption signal is expected from the analytical estimation. It therefore calls for more future observations on UHJs to populate this plot. Detailed modeling of individual planets will be valuable for constraining the hydrodynamic outflows and mass loss rate as traced by atomic absorption lines in the optical.

\section{Conclusion}

With the purpose of detailed characterisation of the ultra-hot Jupiter \M, we carried out transit photometry and radial velocity measurements using EulerCam and CORALIE at the 1.2~m Euler telescope, delivering a refined planet mass of $1.675\pm0.241$ $M_{\rm Jup}$, together with other updated system and planet parameters.
We analysed the optical transmission spectrum of \M observed with the high-resolution spectrograph ESPRESSO at the VLT and report the detection of various species in the atmosphere, including \ion{H}{i}, \ion{Na}{i}, \ion{Mg}{i}, \ion{Ca}{i}, \ion{Ca}{ii}, \ion{Cr}{i}, \ion{Fe}{i}, and \ion{Fe}{ii}. This adds \M to the ensemble of UHJs showing a profusion of atomic absorption features. Putting the measurements into perspective, we explored the trends of atomic absorption features within the UHJ population, indicating two distinct atmospheric regimes as probed through different absorption signatures.
The absorption by metals such as \ion{Mg}{i} and \ion{Fe}{i} appears to  trace the hydrostatic region of atmospheres as the line strengths correlate well with the scale heights of different planets. The H$\alpha$ and \ion{Fe}{ii} absorption strengths, which deviate from the scale height correlation, yet show a positive relation with each other among the UHJ population. Through analytical estimations, we suggest that the correlation is consistent with the exospheric origin of \ion{Fe}{ii} and H$\alpha$ absorption in UHJs, driven by the dominant outflows subject to stellar high-energy radiation. This shows the potential of using both species as probes for the hydrodynamic escape and mass loss of UHJs. Studying the diverse atomic transmission signatures allows us to disentangle the hydrostatic and the exospheric regime of the extremely irradiated planets.

\begin{acknowledgements}
Based on observations collected at the European Southern Observatory under ESO programme 0104.C-0605. We thank the referee for insightful comments that help improve the manuscript. We thank Aline Vidotto for the discussion on the atmospheric escape of ultra-hot Jupiters.
Y.Z. and I.S. acknowledge funding from the European Research Council (ERC) under the European Union's Horizon 2020 research and innovation program under grant agreement No. 694513.
M.L. acknowledges support of the Swiss National Science Foundation under grant number PCEFP2\_194576. The contribution of M.L. and A.P. have been carried out within the framework of the NCCR PlanetS supported by the Swiss National Science Foundation.
DACE is a web platform hosted in Geneva and developed by the Swiss National Center of Competence in Research (NCCR) PlanetS.
\end{acknowledgements}

\bibliographystyle{aa} 
\bibliography{ref} 

\begin{appendix} 
\section{Analytical estimation of atomic absorption in exospheres} \label{app:estimation}

We consider the gas composed of atomic hydrogen in the upper atmosphere (exosphere) escaping the planet with a velocity of $u$ and a constant mass loss rate $\dot{M}$. The conservation of mass provides
\begin{equation} \label{eq:mass_loss_conserve}
    \dot{M} = 4\pi R_c^2 u \rho,
\end{equation}
where $R_c$ is the size of the exobase, $\rho$ is the density of hydrogen, and $u=x_u v_\mathrm{esc}$ is a fraction of the planet's escape velocity $v_\mathrm{esc}=\sqrt{2GM_p/R_p}$.

The energy-limited mass loss rate following \citet{Erkaev2007} is given by
\begin{equation} \label{eq:mass_loss_energy}
     \dot{M} = \frac{\pi R_c^2 F_\mathrm{EUV}\epsilon}{\Phi_0 K},
\end{equation}
where $R_p$ is the planet radius, $F_\mathrm{EUV}$ is the stellar EUV flux at the location of the planet, $\epsilon$ is the heating efficiency, $\Phi_0$ is the gravitational potential at the planetary radius ($\Phi_0=GM_p/R_p$), and $K(\frac{R_{Rl}}{R_p})$ is the coefficient accounting for the potential difference between the Roche lobe boundary ($R_{Rl}$) and the planetary surface ($R_p$) as a result of stellar tidal forces.

Combining Equation~\ref{eq:mass_loss_conserve} and \ref{eq:mass_loss_energy}, we get the number density of H atom in the exosphere as a result of outflows
\begin{equation} \label{eq:density}
    \rho = \frac{\epsilon F_\mathrm{EUV}}{2 x_u K v_\mathrm{esc}^{3}} \simeq \rho_0 \bigg( \frac{\epsilon F_\mathrm{EUV}}{F_0} \bigg) \bigg( \frac{v_0}{v_\mathrm{esc}} \bigg)^3,
\end{equation}
where $F_0=450$ erg cm$^{-2}$ s$^{-1}$, $v_\mathrm{esc}=60$ \kms, and $\rho_0 = 10^{-15}$ g cm$^{-3}$, as informed by previous simulations such as \citet{Murray-Clay2009, Allan2019}.

Equation \ref{eq:density} suggests that under the assumption of the energy-limited mass loss, the exospheric density $\rho$ scales with the planet's escape velocity (or gravitational potential) and the EUV flux received. 

In addition to the outflow, the photoionisation of species also plays a role in determining the level of transmission signal. For a rough estimation of the photoionisation, we simply assume the extended exosphere as a homogeneous optically thin cloud subject to stellar EUV radiation. The characteristic temperature of the exosphere is determined by the balance of heating ($Q$) and cooling ($C$) following \citet{Murray-Clay2009}.
\begin{equation} \label{eq:Q}
   Q =  \epsilon F_\mathrm{EUV} \sigma_{\nu_0} n_n,
\end{equation}
where $\sigma_{\nu_0}$ is the cross section for the photoionisation of hydrogen and $n_n$ is the number density of neutral hydrogen. 
For the cooling term, we assume it is driven by radiative losses resulting from collisional excitation of Ly$\alpha$ line
\begin{equation} \label{eq:C}
    C = 7.5 \times 10^{-19} n_n n_+ \exp[-1.183\times 10^{5}/T],
\end{equation}
where $n_+$ is the number density of protons, equivalent to the number density of electrons. 

Considering the photochemistry of \ion{H}{i}, we solve for the ionisation balance (the rate of photoionisation and radiative recombination) to estimate the degree of ionization $f_\mathrm{H}$.
\begin{equation} \label{eq:ion}
    \frac{F_\mathrm{EUV} \sigma_{\nu_0} n_n}{e_\mathrm{in}} = n_+ n_e \alpha_\mathrm{rec},
\end{equation}
where $\sigma_{\nu_0}=1.89\times10^{-18} \mathrm{cm}^2$ is the cross section for photoionisation of hydrogen \citep{Spitzer1978}; the recombination coefficient $\alpha_\mathrm{rec} = 2.7\times10^{-13}(T/10^{4})^{-0.9}$ taken from \citet{Storey1995}; $n_+ = n_e = n\ f_\mathrm{H}$ and $n_n = n\ (1-f_\mathrm{H})$, where $n=\rho/m_0$, $m_0$ is the mass of H atom; $e_\mathrm{in}$ is the input photon energy, assumed to be 20 eV, and the heating efficiency is $\epsilon = 1-13.6 \mathrm{eV}/e_\mathrm{in}=0.32$. 

Combining Equation~\ref{eq:density}, \ref{eq:Q}, \ref{eq:C}, and \ref{eq:ion}, we solve for the ionization degree of hydrogen and the temperature as follows
\begin{equation} \label{eq:fH}
\begin{aligned}
 \frac{1}{f_\mathrm{H}} &= 1+0.015/(T^{0.9} \exp[-1.183\times 10^{5}/T]), \\
  \frac{1}{f_\mathrm{H}} &= \frac{0.4\rho_0}{m_0 F_0}  \bigg( \frac{v_0}{v_\mathrm{esc}} \bigg)^3 \exp[-1.183\times 10^{5}/T].
\end{aligned}
\end{equation}

Similarly, for other species such as \ion{Fe}{i}, the ionisation balance of Fe combined with Equation \ref{eq:ion} gives
\begin{equation} \label{eq:fFe}
    \frac{1}{f_\mathrm{Fe}} = 1+ \frac{1-f_\mathrm{H}}{f_\mathrm{H}} \frac{\sigma_\mathrm{H}\alpha_\mathrm{rec,Fe}}{\sigma_\mathrm{Fe} \alpha_\mathrm{rec,H}},
\end{equation}
where $\sigma_\mathrm{Fe}=3.66\times10^{-18} \mathrm{cm}^2$ \citep{Verner1996}, the recombination coefficient $\alpha_\mathrm{rec,Fe} = 3.7\times10^{-12}(T/300)^{-0.65}$ \citep{Woodall2007}.

\begin{figure}
  \centering
  \includegraphics[width=\hsize]{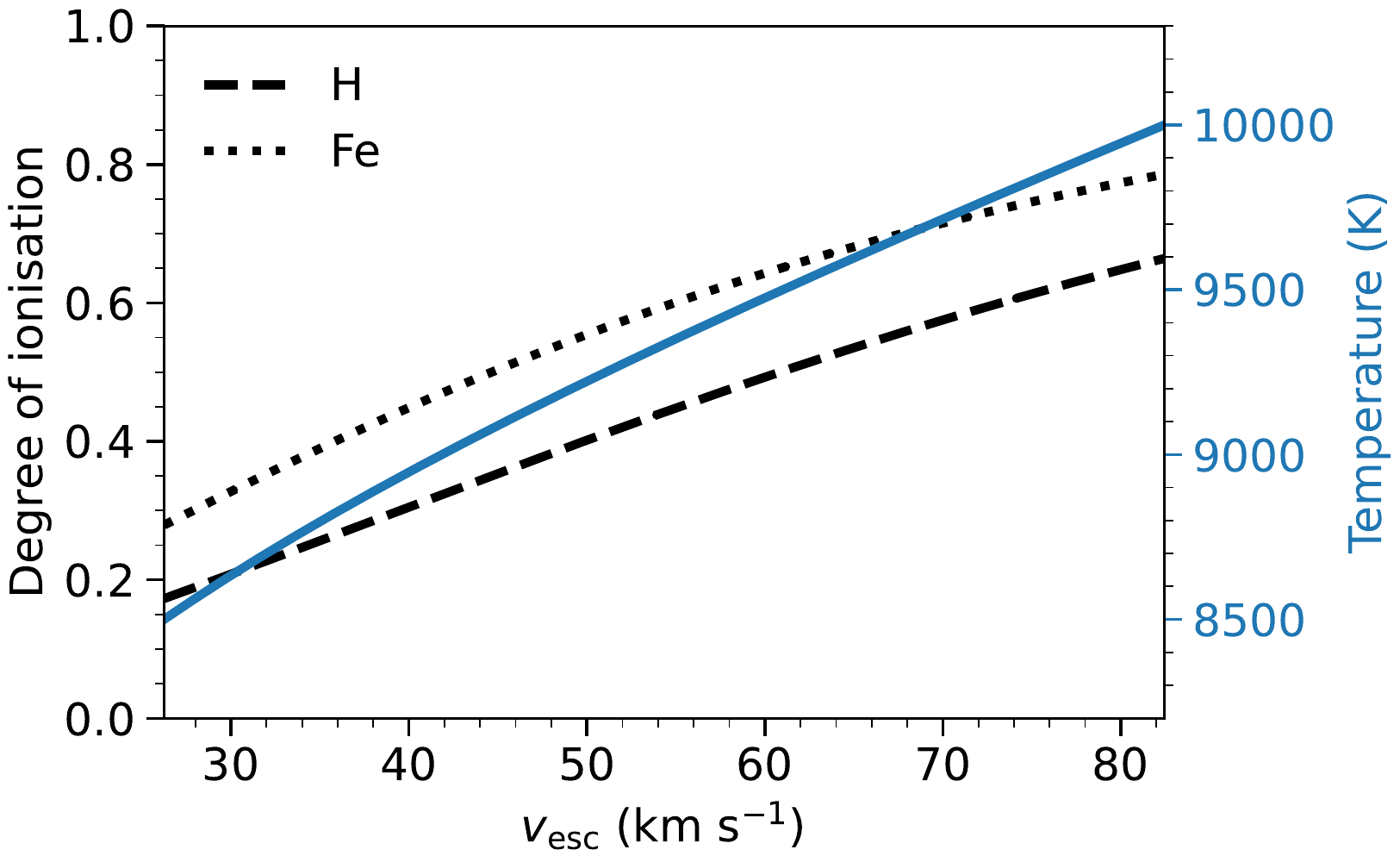}
      \caption{Estimated temperature and degree of ionization of H and Fe for different planets with a range of $v_\mathrm{esc}$. The ionisation fractions are not expected to vary drastically across different UHJs.
              }
         \label{fig:degree_of_ion}
  \end{figure}

Hence, the exospheric temperature and degree of ionisation depend on the planet's potential in terms of $v_\mathrm{esc}$ in our simplified model, as shown in Fig.~\ref{fig:degree_of_ion}.

In the optically thin limit, the equivalent width of absorption is proportional to the total mass of the absorbing material \citep{Hoeijmakers2020a, Gebek2020}. For neutral hydrogen and ionised iron, they can be written as follows
\begin{equation} \label{eq:M}
\begin{aligned}
  \mathcal{T}_\mathrm{H} &\sim \rho (1-f_\mathrm{H}) \sim \rho_0 \bigg( \frac{\epsilon F_\mathrm{EUV}}{F_0} \bigg) \bigg( \frac{v_0}{v_\mathrm{esc}} \bigg)^3 (1-f_\mathrm{H}), \\
 \mathcal{T}_\mathrm{Fe+} &\sim \rho_\mathrm{Fe}  f_\mathrm{Fe} \sim \rho_0 \bigg( \frac{\epsilon F_\mathrm{EUV}}{F_0} \bigg) \bigg( \frac{v_0}{v_\mathrm{esc}} \bigg)^3 A_\mathrm{Fe} f_\mathrm{Fe},
 \end{aligned}
\end{equation}
where a constant mixing of the metal species is assumed, $A_\mathrm{Fe}$ is the mass fraction of iron. 
We note that the photoionisation term ($f_\mathrm{H}$ and $f_\mathrm{Fe}$) can result in variations of the signal strength from different planets by a factor of 4 at most (see Fig.~\ref{fig:degree_of_ion}).
Whereas, equation~\ref{eq:M} contains the linear term of $F_\mathrm{EUV}$ that usually varies by orders of magnitudes from system to system. Therefore, the absorption signal from the exosphere is dominated by the planetary outflow $\dot{M}$ induced by the EUV flux, which drives the positive correlation between H$\alpha$ and \ion{Fe}{ii} as we note in Section~\ref{sec:correlation}.  
Hence such correlation is indeed expected, which reflects properties such as the mass-loss rate and EUV irradiation of different UHJs.

For simplicity we did not take into account the fraction of atoms at the right state (e.g. the principal quantum number n=2 for H$\alpha$ absorption). This requires non-local thermal equilibrium (NLTE) calculation of the radiation field and gets even more complicated for the case of \ion{Fe}{ii} where we combined multiple lines in the observation. The ionisation of \ion{Fe}{ii} is also ignored here. We do not attempt to make detailed interpretation of observed lines in any particular system. Instead, our aim is to draw the relation between H$\alpha$ and \ion{Fe}{ii} across the UHJ population. These assumptions hopefully do not deter the overall trend, yet we need to rely on detailed simulations for a definite answer.

  \begin{figure}
  \centering
  \includegraphics[width=\hsize]{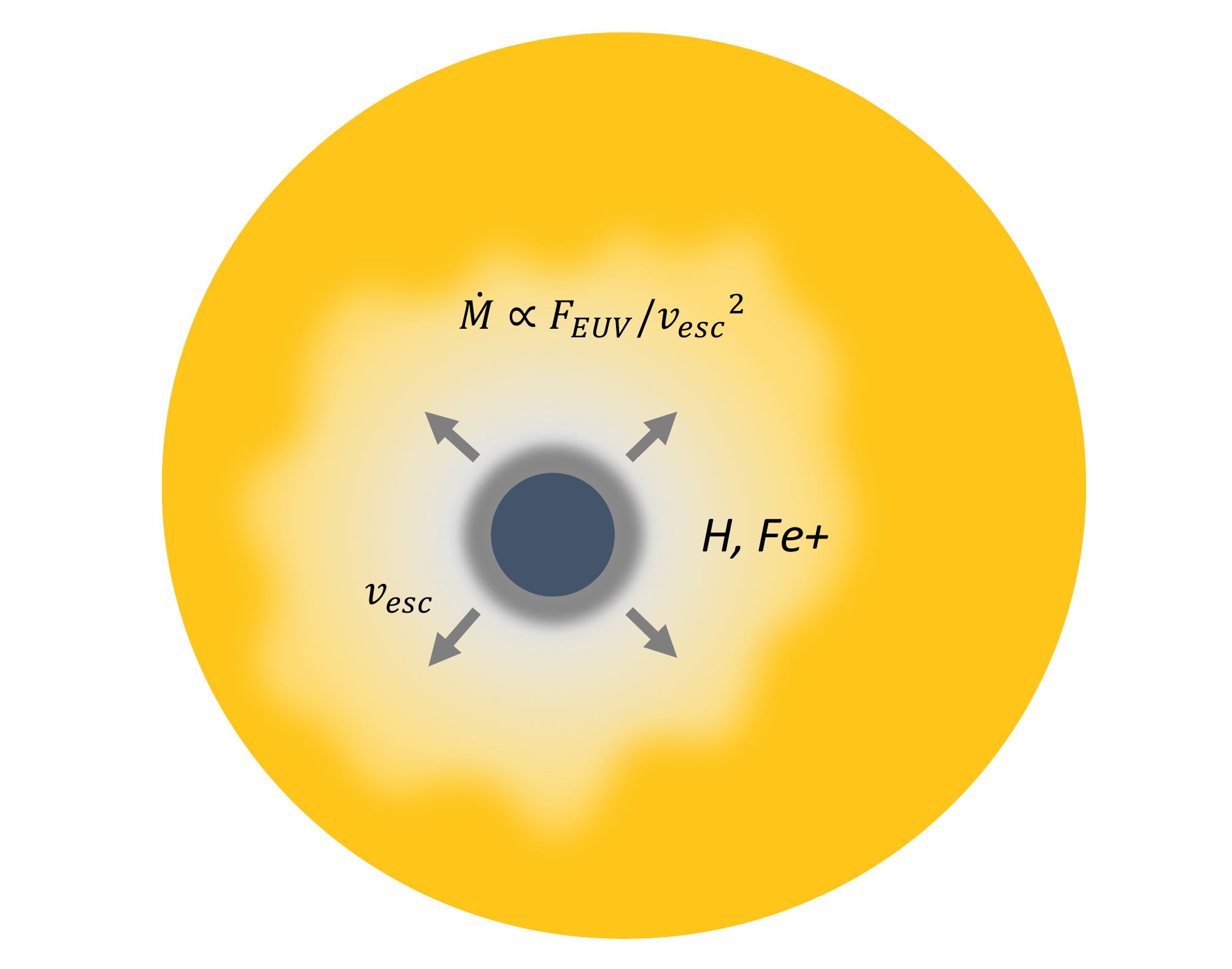}
      \caption{Schematic of a transiting ultra-hot Jupiter with a hydrostatic lower atmosphere and a optically-thin escaping exosphere. 
              }
         \label{fig:shcematic}
  \end{figure}

\section{Summary of UHJs with detailed characterisation with high-resolution transmission spectroscopy}
\begin{table*}
\caption{Properties of atomic absorption features detected in transmission spectra of ultra-hot Jupiters.}
\label{tab:UHJs}      
\centering
\resizebox{\textwidth}{!}{
\begin{tabular}{l c c c c c c c}  
\hline\hline         
& HAT-P-70b & KELT-9b & MASCARA-2b & MASCARA-4b & WASP-76b & WASP-121b & WASP-189b \\

\hline 
$R_p/R_\ast$ & $0.099$ & $0.082$ & $0.113$ & $0.087$ & $0.109$ & $0.125$ & $0.061$ \\
$T_\mathrm{eq}$ (K) & $ 2562 \pm 52 $ & $ 3921 \pm 182 $ & $ 2260 \pm 50 $ & $ 2250 \pm 62 $ & $ 2228 \pm 120 $ & $ 2358 \pm 52 $ & $ 2641 \pm 31 $ \\
$T_\ast$ (K) & $ 8450 \pm 690 $ & $ 9600 \pm 400 $ & $ 8980 \pm 130 $ & $ 7800 \pm 200 $ & $ 6329 \pm 65 $ & $ 6586 \pm 59 $ & $ 8000 \pm 100 $ \\
$R_p$ ($R_\mathrm{Jup}$) & $ 1.87 \pm 0.15 $ & $ 1.926 \pm 0.047 $ & $ 1.83 \pm 0.07 $ & $ 1.515 \pm 0.044 $ & $ 1.863 \pm 0.083 $ & $ 1.865 \pm 0.044 $ & $ 1.619 \pm 0.021 $ \\
$M_p$ ($M_\mathrm{Jup}$) & $ <6.78 $ & $ 2.88 \pm 0.35 $ & $ <3.5$ & $ 1.675 \pm 0.241 $ & $ 0.894 \pm 0.014 $ & $ 1.183 \pm 0.064 $ & $ 1.99 \pm 0.16 $ \\
$H_0$ (km) & $ >184 $ & $ 704 \pm 95 $ & $ >302$ & $ 430 \pm 66 $ & $ 1206 \pm 102 $ & $ 967 \pm 65 $ & $ 485 \pm 40 $ \\
$h$ H$\alpha$ (\%) & $ 1.560 \pm 0.150 $ & $ 1.150 \pm 0.050 $ & $ 0.765 \pm 0.090 $ & $ -0.317 \pm 0.021 $ & $ <0.470$ & $ 1.700 \pm 0.048 $ & $ 0.13 \pm 0.02 $ \\
FWHM H$\alpha$ (\kms) & $ 34.1 \pm 3.9 $ & $ 51.2 \pm 2.6 $ & $ 20.8 \pm 3.4 $ & $ 31.4 \pm 2.4 $ & - & $ 40.9 \pm 1.7 $ & $ 27.4 \pm 4.6 $ \\
$h$ \ion{Fe}{i} (\%) & $ 0.037 \pm 0.003 $ & $ 0.016 \pm 0.001 $ & $ 0.013 \pm 0.002 $ & $ 0.0150 \pm 0.0001 $ & $ 0.049 \pm 0.003 $ & $ 0.039 \pm 0.002 $ & $ 0.0075 \pm 0.0004 $ \\
FWHM \ion{Fe}{i} (\kms) & $ 9.7 \pm 1.0 $ & $ 20.1 \pm 1.0 $ & $ 12.39 \pm 1.69 $ & $ 22.5 \pm 0.2 $ & $ 8.6 \pm 0.7 $ & $ 15.9 \pm 1.1 $ & $ 15.3 \pm 1.0 $ \\
$h$ \ion{Fe}{ii} (\%) & $ 0.437 \pm 0.017 $ & $ 0.183 \pm 0.004 $ & $ 0.130 \pm 0.011 $ & $ 0.0444 \pm 0.0011 $ & $ <0.063 $ & - & $ 0.023 \pm 0.002 $ \\
FWHM \ion{Fe}{ii} (\kms) & $ 13.7 \pm 0.6 $ & $ 21.8 \pm 0.7 $ & $ 8.54 \pm 0.87 $ & $ 16.5 \pm 0.4 $ & - & - & $ 14.1 \pm 1.2 $ \\
$h$ \ion{Na}{i} D (\%) & $ 0.655 \pm 0.150 $ & $> 0.095 \pm 0.007 $ & $ 0.320 \pm 0.050 $ & $ 0.191 \pm 0.017 $ & $ 0.360 \pm 0.050 $ & $ 0.480 \pm 0.047 $ & $ 0.153 \pm 0.031 $ \\
$h$ \ion{Mg}{i} (\%) & $ 0.131 \pm 0.018 $ & $ 0.056 \pm 0.005 $ & $ 0.062 \pm 0.005 $ & $ 0.0221 \pm 0.0009 $ & $ 0.114 \pm 0.016 $ & $ 0.123 \pm 0.015 $ & $ 0.018 \pm 0.002 $ \\
$h$ \ion{Ca}{ii} (\%) & $ 3.750 \pm 0.370 $ & $ 0.780 \pm 0.040 $ & $ 0.560 \pm 0.050 $ & $ 0.775 \pm 0.082 $ & $ 2.440 \pm 0.340 $ & $ 4.700 \pm 0.180 $ & $ 0.40 \pm 0.05 $ \\
References & (1),(2) & (3)-(10) & (11)-(15) & (16)-(18) & (19)-(24) & (25)-(28) & (29)-(31) \\
\hline
\end{tabular}
}
\tablefoot{For the calculation of the lower atmosphere scale height $H_0$, we assume a mean molecular weight $\mu$ of 2.3. The amplitude of \ion{Na}{i} absorption ($h$ \ion{Na}{i} D) takes the average value of the \ion{Na}{i} doublet at 5891 and 5897 \AA. The $h$ \ion{Ca}{ii} takes the average value of the \ion{Ca}{ii} H \& K lines at 3969 and 3934 \AA. Properties of \ion{Mg}{i}, \ion{Fe}{i}, and \ion{Fe}{ii} are taken from cross-correlation outcomes.
}
\tablebib{ (1) \citet{Zhou2019}, (2) \citet{Bello-Arufe2022},  (3) \citet{Gaudi2017}, (4) \citet{Yan2018}, (5) \citet{Cauley2019}, (6) \citet{Borsa2019}, (7) \citet{Hoeijmakers2019}, (8) \citet{Yan2019}, (9) \citet{Turner2020}, (10) \citet{Wyttenbach2020}, (11) \citet{Talens2018}, (12) \citet{Casasayas-Barris2020}, (13) \citet{Hoeijmakers2020b}, (14) \citet{Nugroho2020}, (15) \citet{Stangret2020}, (16) \citet{Dorval2020}, (17) \citet{Ahlers2020}, (18) this work, (19) \citet{West2016}, (20) \citet{Ehrenreich2020}, (21) \citet{Tabernero2021}, (22) \citet{Seidel2021}, (23) \citet{Casasayas-Barris2021}, (24) \citet{Kesseli2022}, (25) \citet{Delrez2016}, (26) \citet{Cabot2020}, (27) \citet{Hoeijmakers2020a}, (28) \citet{Borsa2021}, (29) \citet{Anderson2018}, (30) \citet{Prinoth2022}, (31) \citet{Stangret2022}.
}
\end{table*}  

\end{appendix}

\end{document}